\documentclass[10pt,conference]{IEEEtran}
\AtBeginDocument{%
  \providecommand\BibTeX{{%
    \normalfont B\kern-0.5em{\scshape i\kern-0.25em b}\kern-0.8em\TeX}}}

\usepackage[linesnumbered,ruled,vlined]{algorithm2e}
\usepackage{amsmath,amssymb,amsfonts}
\usepackage{textcomp}
\usepackage{xcolor}
\usepackage{caption}
\usepackage{subcaption}
\usepackage{multirow}
\usepackage{graphicx}
\usepackage{booktabs}
\usepackage{makecell}
\usepackage{float}
\usepackage{enumitem}
\usepackage{listings}
\usepackage{breakcites}
\usepackage{tcolorbox}
\usepackage{multicol} 
\usepackage{subcaption}
\usepackage{ulem}
\usepackage{wrapfig}
\usepackage{amsmath}
\usepackage{cite}
\usepackage{hyperref}

\begin{document}

\newcommand{\todo}[1]{\textbf{\color{red}{Ravishka: #1} }}

\newcommand{\wei}[1]{\textcolor{blue}{Wei: [#1]}}

\newcommand{\ww}[1]{\textcolor{blue}{Weihang: [#1]}}

\newcommand{\td}[1]{\textcolor{blue}{ToDo: [#1]}}

\newcommand{\zijie}[1]{{\color{purple}[Zijie: #1]}}

\newcommand{\zeqing}[1]{{\textcolor{blue}[Zeqing: #1]}}

\newcommand{\jiajun}[1]{{\color{cyan}[Jiang: #1]}}

\newcommand{\distance}{4pt}
\setlength{\textfloatsep}{\distance}

\newcommand\mycommfont[1]{\small\ttfamily\textcolor{violet}{#1}}
\SetCommentSty{mycommfont}

\lstdefinestyle{Cpp}{ % Define a style for your code snippet, multiple definitions can be made if, for example, you wish to insert multiple code snippets using different programming languages into one document
	%    backgroundcolor=\color{highlight}, % Set the background color for the snippet - useful for highlighting
	language=C++,
	basicstyle=\scriptsize\ttfamily, % The default font size and style of the code
	breakatwhitespace=false, % If true, only allows line breaks at white space
	breaklines=true, % Automatic line breaking (prevents code from protruding outside the box)
	captionpos=b, % Sets the caption position: b for bottom; t for top
	commentstyle=\color[rgb]{0.0, 0.5, 0.69},%\color[rgb]{0,0.6,0}, % Style of comments within the code - dark green courier font
	deletekeywords={}, % If you want to delete any keywords from the current language separate them by commas
	%escapeinside={\%}, % This allows you to escape to LaTeX using the character in the bracket
	escapeinside={<@}{@>},
	firstnumber=1, % Line numbers begin at line 1
	frame=lines, % Frame around the code box, value can be: none, leftline, topline, bottomline, lines, single, shadowbox
	frameround=tttt, % Rounds the corners of the frame for the top left, top right, bottom left and bottom right positions
	keywordstyle={[1]\color{blue!90!black}},
	keywordstyle={[3]\color{red!80!orange}},
	morekeywords={String,int}, % Add any functions no included by default here separated by commas
	numbers=left, % Location of line numbers, can take the values of: none, left, right
	numbersep=-8pt, % Distance of line numbers from the code box
	numberstyle=\tiny\color[rgb]{0.1,0.1,0.1}, % Style used for line numbers
	rulecolor=\color{black}, % Frame border color
	showstringspaces=false, % Don't put marks in string spaces
	showtabs=false, % Display tabs in the code as lines
	stepnumber=1, % The step distance between line numbers, i.e. how often will lines be numbered
	stringstyle=\color[rgb]{0.58,0,0.82},
	tabsize=2, % Number of spaces per tab in the code
	backgroundcolor=\color{white}
}
%\usepackage{algpseudocode} 
%\usepackage[dvipsnames]{xcolor}

% \renewcommand\theFancyVerbLine{\small\arabic{FancyVerbLine}}
% \setminted{fontsize=\footnotesize,
%     xleftmargin=8mm,
%     linenos,
%     frame=lines,
%     breaklines}

%\settopmatter{printacmref=false}

%\documentclass[sigconf,review,anonymous]{acmart}
%\documentclass[10pt,conference]{IEEEtran}

\title{CodeImprove: Program Adaptation for Deep Code Models}
\makeatletter % changes the catcode of @ to 11
\newcommand{\linebreakand}{%
  \end{@IEEEauthorhalign}
  \hfill\mbox{}\par
  \mbox{}\hfill\begin{@IEEEauthorhalign}
}
\makeatother % changes the catcode of @ back to 12

\author{
\IEEEauthorblockN{Ravishka Rathnasuriya$^*$}
\IEEEauthorblockA{University of Texas at Dallas\\
USA\\
ravishka.rathnasuriya@utdallas.edu}

\and

\IEEEauthorblockN{Zijie Zhao$^+$}
\IEEEauthorblockA{University of Pennsylvania\\
USA\\
tez@seas.upenn.edu}

\and

\IEEEauthorblockN{Wei Yang}
\IEEEauthorblockA{University of Texas at Dallas\\
USA\\
wei.yang@utdallas.edu}
}

%\title{CodeImprove: Adapting Program Data to Deep Learning Models Through Program Transformations}
%%%\input{tex/story}
%Codeimprove: Program Adaptation for Deep Code Models
%\input{tex/abstract}
\maketitle
\def\thefootnote{*}\footnotetext{Corresponding author.}\def\thefootnote{\arabic{footnote}}
\def\thefootnote{+}\footnotetext{At the time of submission, Zijie Zhao was serving as an intern at the University of Texas at Dallas, USA.}\def\thefootnote{\arabic{footnote}}
\begin{abstract}
%\textcolor{red}{}

Leveraging deep learning (DL)-based code analysis tools to solve software engineering tasks is becoming increasingly popular. 
%However, these tools face challenges related to performance limitations, scalability issues, and the potential for erroneous results (i.e., mispredictions). 
Code models often suffer performance degradation due to various reasons (e.g., code data shifts).
%\textcolor{blue}{Existing approaches on improving the accuracy of these tools face challenges related to performance limitations, scalability issues, and the potential for erroneous results (i.e., mispredictions).}
Retraining is often required to address these issues, but frequent model updates are costly in labeling and deployment. In this paper, we explore an alternative solution: Adapting the program inputs to the code models. This can be achieved by two steps: 1) input validation that focuses on identifying whether an input is an out-of-scope input program that are beyond a model’s handling capability, and 2) input adaptation that adapts out-of-scope inputs to become in-scope inputs.
%%%Leveraging deep learning (DL) code models to solve software engineering (SE) tasks is becoming increasingly popular. However, code data shifts, often caused by software evolution, degrade the performance of deep code models by creating syntactic mismatches between training and testing code data. To address this issue, retraining is often required, but frequent model updates are costly in labeling and deployment. In this paper, we explore an alternative solution: Adapting the program inputs to the code models. This can be achieved by two steps: 1) input validation that focuses on identifying whether an input is an out-of-scope input program that are beyond a model’s handling capability, and 2) input adaptation that adapts out-of-scope inputs to become in-scope inputs.
%On the one hand, validating input is challenging because the existing techniques are mainly focused on image data and cannot be used for code data due to different data shift patterns (e.g.: code syntax change versus image matrix transformation). On the other hand, adapting an out-of-scope program can be challenging due to large search spaces, i.e., an out-of-scope program can be modified in a significantly large number of ways while preserving the program semantics. Designing an input validation approach specifically for code data for input validation and applying a set of program transformations to limit the search space can solve the aforementioned challenges. 
Validating program input is challenging, as current techniques focus on continuous inputs such as image data and fail with discrete inputs like code data, which have unique characteristics and are processed differently by deep learning models. Adapting out-of-scope programs is also challenging due to their vast search spaces.
Therefore, in this paper, we propose CodeImprove, which distinguishes out-of-scope from normal inputs and converts such out-of-scope inputs back to in-scope inputs through program transformation. In particular,  we propose a validity score metric to identify out-of-scope inputs and leverage genetics algorithms to apply semantic preserving program transformation to convert out-of-scope inputs to in-scope inputs. Our experimental results show CodeImprove can enhance upto 8.78\% of accuracy, and 51.28\% of relative improvements in three code models on two SE tasks. Additionally, our input validation is promising in detecting out-of-scope inputs (AUC score of 0.924).

\end{abstract}
\section{Introduction}

In the field of software engineering, code analysis tools play a crucial role in addressing critical tasks such as vulnerability detection, defect prediction, and clone detection~\cite{tian2023fly, naturalattack, Zhang2023Challenging, lu2021codexglue}. 
%\textcolor{blue}{DL models approximate solutions based on learned patterns in training datasets, which may not be representative of all the possible data. Code models often suffer performance degradation due to various reasons~\cite{yang2024robustness, yefet2020adversarial,hu2023codes,li2021estimating}.}
Code analysis tools often suffer performance degradation due to various reasons~\cite{yang2024robustness, yefet2020adversarial,hu2023codes,li2021estimating,van2020tailoring,peng2018t}.
%\textcolor{blue}{However, existing approaches~\cite{yuDataAugmentationProgram2022,xiao2021selfchecking,xiao2022repairing} on improving the accuracy of these tools face challenges related to performance limitations, scalability issues, and the potential for erroneous results.}
%These tools often face challenges related to performance limitations, scalability issues, and the potential for erroneous results. 
Traditionally, the primary approach to overcoming these obstacles has been to refine the tools to handle various challenging cases, which might involve updating their versions or incorporating new heuristics. However, this process can be complex and resource-intensive. Recently, researchers have identified an alternative strategy: \textit{adapting the inputs to the tools}~\cite{van2020tailoring,peng2018t}. This approach has gained traction, especially in scenarios where refining the tools becomes impractical. This adaptation of inputs serves as an effective strategy to circumvent the limitations of tool modification, ensuring that code analysis tools remain effective, even for cases they were initially unable to address before the adaptation of inputs.

While existing input adaptation efforts have primarily focused on traditional symbolic reasoning-based tools~\cite{van2020tailoring,peng2018t}, there is a growing need for developing input adaptation strategies for (deep) learning-based code analysis tools/models~\cite{Devlin2019BERT,Sanh2020DistilBERT,guo2021graphcodebert,Chen2021Evaluating}. This need arises from the challenges and high costs associated with improving these models to address their limitations. Commonly, improving these tools involves retraining and replacing the underlying deep learning models~\cite{yuDataAugmentationProgram2022,xiao2021selfchecking,xiao2022repairing}. Such processes not only lead to increased labeling and computing efforts but also risk reducing the models' generalization capabilities, alongside potential compatibility and version control issues. Additionally, model replacement may need the creation of new architectures and datasets, which incurs substantial costs related to rebuilding and redeploying the systems.

To address the challenges, an alternate cost-effective solution is to adapt the program inputs to learning-based tools without altering the original tool. This strategy of input adaptation is particularly beneficial in scenarios like agile development, where the code base rapidly changes, as it maintains the model's applicability without the need for frequent retraining. Additionally, exploring different input adaptation techniques is more resource-efficient than the continuous retraining of a model to discover the optimal approach. 

The process of adapting an input to a code model generally involves two key steps: (1) \textit{input validation} that aims at identifying out-of-scope inputs that fall outside the model's capacity (i.e., inputs prone to being mishandled) , and (2) \textit{input adaptation}, where the out-of-scope inputs are converted by semantic-preserving transformations to become in-scope inputs that are within the model's handling capabilities (i.e., inputs that the model is likely to process correctly).

The process of validating and adapting inputs for code models encompasses distinct challenges. First, in terms of input validation, it is difficult to create a metric that accurately predicts the likelihood of a model generating a correct or incorrect output for a given input. The machine learning community has developed some methodologies known as uncertainty metrics~\cite{guo2017calibration,wang2020dissector,hendrycks2018baseline,gal2016dropout, alon2019code2vec,xiao2019quantifying,vasudevan2019towards,corbiere2019addressing, monarch2021human, steinhardt2016unsupervised,shannon1948mathematical} to estimate a model's level of uncertainty for a specific input.
However, these metrics, which are primarily designed for image data, often fall short when applied to code data in identifying out-of-scope inputs. The root of this issue lies in the nature of image data, which is typically continuous and dense with fewer semantic nuances, allowing for smooth gradients that make the uncertainty metrics more reliable. In contrast, code data is inherently discrete, structured, and filled with abstractions such as control flows and data dependencies, which can result in abrupt changes in the model's output and make the uncertainty less predictable. Moreover, while image data mainly undergoes noise, corruption, or compression as matrix transformations, code data faces changes in syntax and programming paradigms. Our preliminary study (Section~\ref{study}) suggests that these essential distinctions pose a substantial challenge to the direct application of standard uncertainty metrics to code data.

%Notably, our preliminary highlights several insights. First, on average the uncertainty metrics perform 0.42-0.59 in AUC score for two SE tasks (i.e., vulnerability detection and defect prediction) across three deep code models (i.e., CodeBERT, RoBERTa, and GraphCodeBERT) in detecting out-of-scope inputs. We find that the reason for low AUC scores is that the effectiveness of statistical analysis in these metrics. Mainly, these metrics undergo computations such as softmax, entropy etc. However, out-of-scope data may behave similar to that of in-scope data resulting an overlapping situation. Second, disparity in model's behavior on code data. Uncertainty metrics like vanilla show strong performance in RoBERTa model, however, this performance degrades in GraphCodeBERT model. Third, uncertainty metrics are not efficient in detecting out-of-scope data within the same distribution. 
Second, once specific inputs have been identified for adaptation, transforming these inputs into the model's handling scope is also challenging. First, navigating the vast search spaces involved in code transformation is a complex task. Considering the numerous possible transformations that can be applied to code and the fact that many of these transformations can be applied repeatedly to produce various forms of code, the resulting array of potential variations creates a complex landscape for exploration. Moreover, any modifications made during the program transformation process must maintain the original program functionality and semantics of the code. Therefore, efforts to modify the code to enhance model compatibility must be carefully balanced to avoid unintentionally changing its fundamental meaning or functionality, thus posing a dilemma between improving model performance and preserving code semantics.

To address these challenges, we
propose CodeImprove (Figure~\ref{fig:Overview}), the first techniques for input validation and input adaptation of code inputs. 
For input validation, we identified that existing uncertainty metrics misrepresent the model's handling capability on code inputs, leading to overconfident predictions for out-of-scope code inputs (Section~\ref{study}). We observe that the relevance of different aspects of the input, such as structural information or variable names, can shift dynamically across the model's layers. Traditional uncertainty metrics, which typically focus on the outputs of the final few layers, fail to capture this layer-by-layer processing.

Based on such observation, we propose a Dropout-based Sub-Model Generation (DSMG) approach to find an optimal hidden state representation that accurately identifies in-scope versus out-of-scope inputs. By analyzing sub-models derived from the original DL model, CodeImprove can delve into how inputs are processed at each layer. DSMG allows CodeImprove to generate sub-models that provide deeper insights into the transformation of inputs through the network. CodeImprove utilizes the confidence levels of these sub-models' predictions as a new metric for assessing the validity of inputs, offering a more reliable measure that captures the complexities of code input processing in DL models.

% First, we identify out-of-scope inputs from in-scope inputs. %To achieve this, we employ a guiding metric. 
% Although the code inputs to the model can come the same distribution, the uncertainty metrics which performs better at detecting out-of-distribution data~\cite{li2021estimating}  fails to capture out-of-scope inputs from the same distribution with different syntax. Therefore, identifying the most influential code semantics that affects the model's prediction is important. As a solution, we designed our guiding metrics using sub models that represents different layers of the code model. These sub-models helps to perform a granular analysis of different features of the input data and provide a guiding metric by assessing the impact of each layer on decision making. Our objective is to assign higher validation scores to correctly-predicted inputs that indicates a strong prediction while lowering the validation scores for mispredicted inputs. We focus on adjusting validation scores to either increase or decrease  based on sub model's prediction. This objective monitors a code model's inputs and isolate those possibly beyond its model’s handling capability from
% impacting the model’s decision-making.

%Unlike the existing approach~\cite{wang2020dissector} that confuses the correct and incorrect predictions, we focus on adjusting validity scores to either increase or decrease them based on sub model's prediction. 

%Therefore, we identify out-of-scope inputs from in-scope inputs using sub-models that represents different layers of the DL model. 

Following input validation, CodeImprove employs Adaptation by Evolutionary Search (AES). 
We develop a list of basic semantic preserving transformations and leverage DSMG’s validation score as a guiding metric  to combine these basic transformations into a
composite transformation that effectively covert the input
from being out-of-scope to in-scope. 

We evaluated our technique with pre-trained transformer-based language models on software engineering tasks such as vulnerability detection and defect prediction. Our experimental results report promising results and show CodeImprove can enhance 8.78\% of absolute accuracy, and 51.28\% of relative improvements in three code models on two code tasks. Notably, our validity score computation that validates out-of-scope inputs obtained promising results (AUC score of 0.924). 

We summarize our contributions in this paper below:
\begin{itemize}[topsep=2pt]
    \item \textbf{Novel Perspective}. We propose a novel perspective of differentiating out-of-scope from in-scope inputs, as well as adapting these out-of-scope inputs to become in-scope inputs. To the best of our knowledge, our novel perspective is the first attempt to adapt inference-time inputs for deep code models through program transformation. 
    \item \textbf{Tool Implementation}. We implement CodeImprove following the novel perspective (1) by implementing a sub-model generation technique from the original code model for code data, (2) designing a validity score metric to distinguish out-of-scope inputs from in-scope inputs utilizing the generated sub models, and (3) designing a genetic algorithm based technique to adapt out-of-scope inputs to become in-scope inputs by applying program transformation.
    \item \textbf{Comprehensive Evaluation}. We conducted an extensive study on three popular pre-trained models and two code-base tasks, demonstrating the effectiveness and efficiency of CodeImprove's input validation and input adaptation on test data. 
    \item \textbf{Public Artifact}. We release all experimental data and source code at the project Github repository~\cite{Data} for future research, practical use, and experiment replication.   
\end{itemize}

\section{Background}
\label{background}

%%%%ISSTAAA \textcolor{red}{Organize according ly. Problem \texttt{->} there are two sub problems \texttt{->} 1) input valudation, 2) how to adapt the inputs. Show different ways in the following sections.  Input validation-> we have adifferent ways to search for the metric. however, the challenge is if we can detect it with minimum false positive rates. As solution, we have uncertainty, cross-validation, multi-model etc. From here uncertainty is the widely used approach to detect mispredicted inputs. Input adaptation, Once we find the inputs to fix, our next challenge is to how to implement a search a strategy to understand the most influential part of code. This can be achieved via search based testing. There are many techniques employed for this task. Eg: random search, hill climbing, A*, genetics algorithms. Now we can leverage these search techniques to find the code variants to transform without breaking semantics. Then we can utilize our guiding metric to determine which input is the best candidate to adaptation.}

%ISSTA In this section we review uncertainty in DL, genetic algorithms and program transformations, which are key to our solution.

% \jiajun{In general, the intention of the background is not so clear. Some of them are well-known concepts, such as the phases in GA. Maybe, more detailed information that will be used in the subsequent presentation should be introduced. For example, a formalization of the working process?}

\subsection{Problem Definition}

Given a code model \( M \) and an input code snippet \( x \), the class with the highest probability is the final prediction result of \( M \) for \( x \), denoted as \(y = M(x) \). During deployment, ensuring the correctness of every prediction is challenging. Thus, the objective is to enhance model performance on test inputs through code adaptation.

The validation metric \(V(M,x)\)  evaluates \(M\)'s uncertainty on input \(x\) to determine whether the input is in-scope or out-of-scope. If \(V(M,x)\) is less than the predefined threshold \(c\), it indicates uncertainty, and \(x\) requires adaptation. Otherwise, the \(x\) is considered in-scope. The set of transformations \(T = \{T_1, T_2, \ldots, T_n\}\) refers to a sequence of code transformation operators applied to the out-of-scope input \(x\), resulting in a modified input \(x'\). Let \( \hat y \) represent the ground truth of \( x \) and let \( y' = M(x') \) be the prediction result after adapting \( x \) to \( x' \) via  \( T \).

The goal is to compute \( V \) and apply \( T \) such that the loss function of the adapted prediction \( L(y') \) is smaller than the original loss \( L(y) \). We aim to find \( V \) and \( T \) to make \( L(y') < L(y) \) where:
\vspace{-1mm}
\begin{equation*}
y' = 
\begin{cases} 
M(T_1, T_2, \ldots, T_n(x)), & \text{if } V(M,x) \leq c, \\ 
y, & \text{if } V(M,x) > c,
\end{cases}
\end{equation*}
% \begin{flushright} 
% \text{where \( c \) is a threshold.}
% \end{flushright}}

The loss \( L(y) \) is characterized by the distance from the ground truth \(\hat{y} \): 
{\small\begin{equation*}
L(y) = \|y - \hat{y}\|
\end{equation*}} 

This can be generalized to any distance metric (e.g., \( L_1 \),\( L_2 \), \(L_{\infty}\))~\cite{carlini2017towards, papernot2016distillation, szegedy2013intriguing, papernot2016effectiveness} to accommodate for any SE task. 

%%edited \textcolor{blue}{This formulation represents the \( L_n \), allowing the use of any distance measure such as \( L_1 \) or \( L_2 \), making it suitable for any SE task.}

%\textcolor{blue}{The loss \( L \) is characterized by the distance from the ground truth:}

% \begin{equation*}
% y' = M(T_1, T_2, \ldots, T_n(x)), \text{ if } V(M(x)) < c, 
% \end{equation*}

% and 

% \begin{equation*}
% y' = \hat y, \text{ if } V(M(x)) > c, 
% \end{equation*}
% \noindent where \( c \) is a threshold

%%%edited one
% \textcolor{blue}{\begin{enumerate}
%     \item \( y = M(x) \)
%     \item If \( V(M(x)) \leq c \), where \( c \) is a threshold, then:
%     \begin{enumerate}
%         \item Apply transformations \( T_n \) to obtain \( x' \):
%         \[
%         x' = T_1, T_2, \ldots, T_n(x)
%         \]
%         \item Predict the new input \( x' \) with \( M \) to obtain \( y' \):
%         \[
%         y' = M(x')
%         \]
%         \item Ensure the loss of \( y' \) such that the new loss \( L(y') \) is smaller than the original loss \( L(y) \):
%         \[
%         L(y') < L(y)
%         \]
%     \end{enumerate}
% \end{enumerate} }

% \textcolor{blue}{The loss \( L \) is characterized by the distance from the ground truth:}% For example, a simple loss function could be:}

% \textcolor{blue}{\[
% L(y') = 
% \begin{cases} 
% 0, & \text{if } y' = y \\
% 1, & \text{if } y' \neq y
% \end{cases}
% \]}
%%edited one end

The challenge is to develop an effective validation metric \( V \) (Oracle problem) and a determine a sequence of transformations \( T \) (Search problem) that adapt out-of-scope inputs.% and improve the model's performance without retraining. }

%Given a code model $M$ and an input code snippet $x$, the class with the largest probability is the final prediction result of $M$ for $x$, denoted as $M(x)$. If $M(x)$ is different from the ground truth label of $x$, it \textcolor{blue}{indicates} that $M$ \textcolor{blue}{has made} a misprediction on $x$; otherwise \textcolor{blue}{it indicates} a correct prediction. However, \textcolor{blue}{$M$ does} not guarantee the correct prediction for each test input during the deployment. This work aims to improve the model performance on test inputs via code adaptation, which makes it an effective technique different from others that require retraining \textcolor{blue}{or replacing the model}~\cite{yuDataAugmentationProgram2022, xiao2021selfchecking,xiao2022repairing}. 

%The process of input adaptation for code data possesses two challenges: 1) \textbf{Oracle problem:} validating a given test input as in-scope or out-of-scope, and 2) \textbf{Search problem: } how to adapt the out-of-scope input to become in-scope input. \textcolor{blue}{Due to space limitations, the formal definitions can be found on our project website~\cite{CodeImprove}.}

\subsection{Oracle Problem- Developing V}
During deployment, determining whether a prediction is correct without manual analysis is challenging. An effective validation metric \(V\) is needed to automatically guide \(M\) to accurately make decisions, thus reducing false positives.  A substantial progress has been made in this direction such as handling uncertainty~\cite{guo2017calibration,wang2020dissector,hendrycks2018baseline,gal2016dropout, alon2019code2vec,xiao2019quantifying,vasudevan2019towards,corbiere2019addressing, monarch2021human, steinhardt2016unsupervised,shannon1948mathematical}, deep emsemble~\cite{lakshminarayanan2017simple}, and cross-layer dissection~\cite{wang2020dissector}.

%%%edited During the deployment, \textcolor{blue}{it is not feasible to precisely determine whether the prediction for a given code input is correct without manual analysis.} 
%we cannot exactly determine whether the prediction for a given code input is correct or not without manual analysis.
%edited This means that it is \textcolor{blue}{challenging} to automatically decide whether a given code input is correctly predicted or not. \textcolor{blue}{To} automate this process, it is crucial to design an effective metric that can guide these code models to accurately make decisions (i.e., reducing false positives). A substantial progress has been made in this direction such as handling uncertainty\textcolor{blue}{~\cite{guo2017calibration,wang2020dissector,hendrycks2018baseline,gal2016dropout, alon2019code2vec,xiao2019quantifying,vasudevan2019towards,corbiere2019addressing, monarch2021human, steinhardt2016unsupervised,shannon1948mathematical}}, deep emsemble~\cite{lakshminarayanan2017simple}, and cross-layer dissection~\cite{wang2020dissector}. %Although these 
%~\cite{guo2017calibration,wang2020dissector,hendrycks2018baseline, li2021estimating, alon2019code2vec,} 

%\textcolor{blue}{~\cite{guo2017calibration,wang2020dissector,hendrycks2018baseline, li2021estimating, alon2019code2vec,monarch2021human, steinhardt2016unsupervised,shannon1948mathematical}}

\subsection{Search Problem- Transformation Sequence T}
Beyond validation, the process of adapting the out-of-scope inputs to become in-scope inputs necessitates efficient search algorithms to explore program syntax for potential transformations. The goal of search techniques is to optimize the code transformations while preserving the program semantics. Search techniques like random search~\cite{zabinsky2009random}, hill climbing search~\cite{selman2006hill}, and genetics algorithms~\cite{harman2001search} offer solutions for code transformations. 
%The variants of search techniques, such as random search~\cite{zabinsky2009random}, hill climbing search~\cite{selman2006hill}, and genetics algorithms~\cite{harman2001search} offer solutions to code transformations for search. 
The search strategy begins with a set of candidate solution(s) generated by applying semantic preserving code transformation. 
These candidates are then evaluated using \(V\) to select the most promising one.
%Then, these candidate solutions are guided by the metrics in the validation phase to select the most promising candidate. 
The search algorithm iteratively refines the candidates until a termination criterion is reached or an optimal solution is found.

\section{Preliminary Study on Input Validation for Code Models}
\label{study}

In our preliminary study,  we assess the applicability of existing uncertainty metrics within the realm of code models, aiming to identify a dependable threshold score that can differentiate in-scope and out-of-scope program inputs. We perform the study on a comprehensive set of metrics~\cite{guo2017calibration,wang2020dissector,hendrycks2018baseline, li2021estimating, alon2019code2vec} from the existing literature. 
%In particular, we calculate an uncertainty score for each test input to find the best threshold to distinguish between in-scope and out-of-scope program inputs.
Our goal is to answer the research question: How effective are the existing uncertainty metrics in distinguishing in/out-of-scope program inputs? 

\subsection{Experimental Method and Setup}

%In order to comprehensively capture the uncertainty in the predictions of DL models, we utilize a selection of well-regarded methods from the existing literature based on their prevalence, scalability, and practical applicability in a real-world scenario.
%as shown in Table ~\ref{table:uncertainty_metrics}. 

\textbf{Uncertainty Metrics: } 
Our study includes eight different uncertainty metrics. We utilize vanilla~\cite{hendrycks2018baseline} that computes maximum softmax probability as the confidence. Temp Scale~\cite{guo2017calibration} is a post-hoc calibrated confidence metric applied to the validation set, where the BFGS optimizer~\cite{international1990bfgs} is used to train a calibration temperature with a learning rate of 0.01. Our study includes confidence-based uncertainty metrics, such as least confidence~\cite{monarch2021human}, which calculates the difference between 100\% confidence and the most confidently predicted label; margin confidence~\cite{monarch2021human}, which determines the difference between the top two most confident softmax predictions; and ratio confidence~\cite{monarch2021human}, which computes the ratio between the top two most confident softmax predictions. We also include uncertainty metrics that were designed using information theory~\cite{steinhardt2016unsupervised,shannon1948mathematical}. Entropy computes the average amount of surprise/ uncertainty for a given outcome. Predictive entropy quantifies the uncertainty associated with the outcomes for a given set of observed inputs. Mutual information measures the amount of information obtained from one random variable given another using entropy and conditional entropy. 
Monte-Carlo Dropout (MCD)~\cite{gal2016dropout} quantifies uncertainty by averaging the logits over multiple dropout samples. Deep Ensembles (DE)~\cite{lakshminarayanan2017simple} quantifies uncertainty by averaging the outputs from multiple independently trained models with different initial seeds. Both employ the sampled winning score (SWS) as the primary uncertainty metric.

\textbf{Dataset and Models: } To evaluate the effectiveness of detecting out-of-scope data through uncertainty quantification, we consider two code tasks and two associated datasets (i.e., defect prediction with CodeChef dataset and vulnerability detection on Devign dataset) on three pre-trained models. More information on datasets and subject models is explained in Section~\ref{subjects}.
\begin{table}[htb!]
\caption{AUC Comparison on Distinguishing Out-of-Scope Inputs for Selected Uncertainty Metrics}
\label{Tab:preliminary}

\resizebox{\columnwidth}{!}{
\begin{tabular}{|c|c|c|c|c|c|c|c|}
\hline
 \textbf{Experiment} &\multicolumn{3}{c|}{\textbf{Vulnerability Detection}} & \multicolumn{3}{c|}{\textbf{Defect Prediction}}\\ \cline{2-7}
      & CodeBERT & RoBERTa & GraphCodeBERT  & CodeBERT & RoBERTa & GraphCodeBERT \\ \hline

    Vanilla  & 0.552 & 0.574  & 0.455  & 0.595 & 0.580 & 0.494   \\\hline

     Temp. Scaling &0.420 & 0.572  & 0.442  & 0.403 & 0.413 & 0.493 \\\hline

     Predicitive Entropy & 0.582 & 0.454  & 0.508  & 0.573 & 0.579 & 0.491\\\hline

    Entropy & 0.414 & 0.556  & 0.482  & 0.436 & 0.391  & 0.426 \\\hline

      Mutual Information & 0.586 & 0.443  & 0.531  & 0.566 & 0.608 & 0.579\\\hline

     Least Confidence & 0.589 & 0.452  & 0.558  & 0.595 & 0.593 & 0.508\\\hline

     Ratio Confidence & 0.464 & 0.521  & 0.486  & 0.553 & 0.484 & 0.469 \\\hline

     Margin Confidence & 0.473 & 0.521  & 0.489  & 0.562 &0.532 & 0.470\\\hline

    % Dissector  & 0.731 & 0.820  & 0.798  & 0.889 & 0.828  &0.875 \\\hline

    MCD & 0.624 & 0.613 & 0.617 & 0.614 & 0.607 &0.616 \\\hline

    DE  & 0.507 & 0.519 & 0.507 & 0.561 & 0.562 & 0.571 \\\hline

\end{tabular}
}
\end{table}

\textbf{Evaluation Metric: }We used the Area Under Curve (AUC)~\cite{davis2006relationship} based on
True Positive Rate (TPR) and False Positive Rate (FPR) data to measure how effective a technique is in distinguishing an out-of-scope inputs from in-scope inputs. AUC quantifies the probability that a positive example receives a higher predictive score than a negative sample. For example, a random classifier yields an AUC of 0.5, while a perfect classifier achieves an AUC of 1.

To compute the AUC scores, we define positive and negative samples based on the correspondence between predicted outputs and ground truth labels. A positive sample indicates correct predictions (in-scope), labeled as 1, while a negative sample signifies misclassified (out-of-scope), labeled as 0. For instance, given a well-trained model $f(\cdot)(·|\theta)$, an input pair $(x,y)$ has ground truth 1 if $f(x|\theta)$ is an exact match of $y$, and 0 otherwise. During evaluation, each uncertainty method predicts a score reflecting the model's capability in handling the input.

\subsection{Results and Analysis: }

Table~\ref{Tab:preliminary} shows AUC scores for the evaluated uncertainty metrics.
The majority of these metrics exhibit AUC scores close to 0.5, akin to what one would expect from a random classifier, with none surpassing an AUC score of 0.624. From these results, it is inferred that the uncertainty metrics under consideration are ineffective at distinguishing between in-scope and out-of-scope code inputs.

\label{design}
\begin{figure}[!htbp]
\centering
\includegraphics[width=\columnwidth]{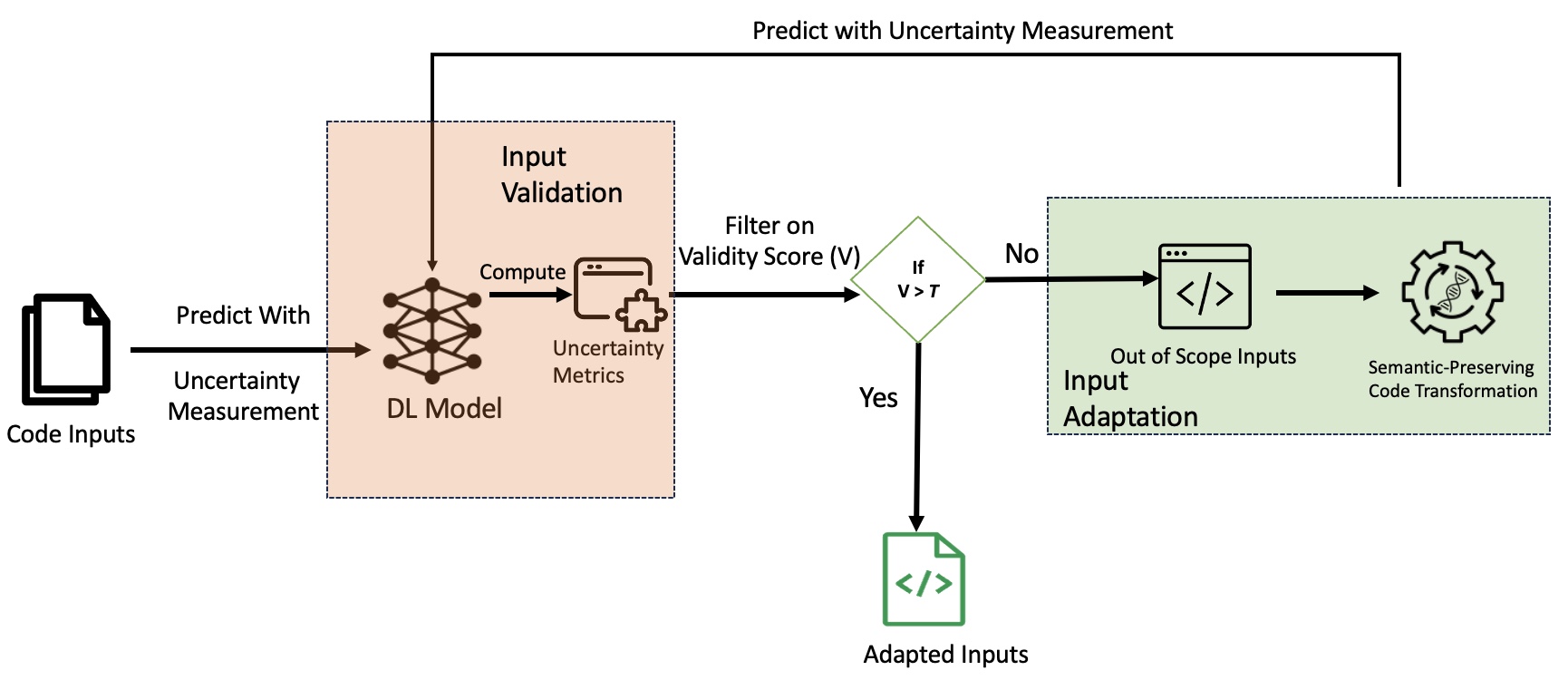}
\caption{Overview of CodeImprove} 
% %\textcolor{red}{TODO: DRAW A FLOW DIAGRAM TO SHOW MANIFESTATION }
\label{fig:Overview}
\end{figure}
\section{Design of CodeImprove}

% Describe overview diagram. 

% Add the genetic algorithms psuedocode. 

% Explain into each section: 
%       -uncertainty measurement to find the threshold. 
%       - using the threshold as the fitness value in program transformation
%       -  Test with new transformed data

% \subsection{Overview}
Figure~\ref{fig:Overview} provides an overview of CodeImprove. 
% Given a trained DL model and test instances, our goal is to develop a systematic method that can adapt out-of-scope inputs into in-scope inputs of DL models.  
% As shown in figure~\ref{fig:Overview}, 
CodeImprove consists of two main technical phases: \textit{Input validation} phase to detect out-of-scope inputs from normal inputs, and \textit{Input adaptation} phase to transform out-of-scope inputs to in-scope inputs. %\textit{Finetuning the DL Model, Identifying in-scope and out-of-scope data and the differentiable threshold, and Transforming out-of-scope inputs}. 

\subsection{Input Validation}
%%%\textcolor{red}{Dropout based Sub-Model Dissection (DSMD) }
%\textcolor{red}{general idea: how things should be added, like GA, mutation step, }
\label{validation}

The goal of this phase is to design a guiding metric that distinguishes between in-scope and out-of-scope test data for a trained DL model and its associated dataset. Our preliminary investigation (Section~\ref{study}) demonstrates that traditional uncertainty metrics are inadequate for software engineering (SE) tasks, motivating the need for a more granular approach to understanding input behavior.

We aim to identify an optimal hidden state representation that distinguishes between in-scope and out-of-scope inputs. Code inputs fed into the DL model's layers often experience shifting focus across various aspects such as from structural information to variable names. Existing uncertainty metrics which rely on outputs from the final few layers fails to capture the this dynamic layerwise processing. To bridge this gap, we explore sub-models extracted from original model to gain deeper insights into the processing of inputs. These sub-models enable us to measure confidence at multiple levels, forming the basis for a more reliable input validation metric. 

A key motivation for our approach is the need to compute epistemic uncertainty, which arises from the model's lack of knowledge about certain inputs and reflects how confident the model is when making predictions. Accurately estimating epistemic uncertainty requires introducing model variance, which captures the variability in predictions when slight changes are applied to the model's architecture or processing. Existing approaches, such as dropout-based techniques~\cite{gal2016dropout} or ensemble methods~\cite{lakshminarayanan2017simple}, aim to create model variance but face significant limitations in the context of code data. These methods do not investigate layerwise processing, ignoring the dynamic transformations that inputs undergo at different model layers, nor do they explore hidden layer representations, which are crucial for understanding how the model processes structured data like code. Most importantly, they are primarily designed for continuous input spaces like images, where smooth gradients and variations can be leveraged for uncertainty estimation. 

%These methods overlook layerwise processing, failing to account for the transformations inputs undergo across different model layers. Additionally, they do not utilize hidden layer representations, which are critical for understanding the structured and discrete nature of code. Furthermore, these methods are primarily designed for continuous input spaces like images, where smooth gradients and variations can be leveraged for uncertainty estimation, making them unreliable for code data.

%%icse major revision We aim to identify an optimal hidden state representation that distinguishes between in-scope and out-of-scope inputs. When code inputs are fed into the DL model, the focus of the model's layers may shift across various aspects of the inputs, such as from structural information to variable names. Existing uncertainty metrics fall short for code inputs because they focus on hidden state outputs of the final few layers of the DL model,  rather than the layer-by-layer processing of inputs. To bridge this gap, we explore sub-models extracted from the original model to gain insights into the processing of inputs. We have devised a method for generating these sub-models and propose using the confidence levels of sub-model predictions as a measure of input validity. 

%\textcolor{red}{Add citation to support. 3.3.1: A survey of uncertainty in deep neural networks.}
%The motivation for using sub-models is driven by the analogy that a group of decision-makers tends to make better decisions than a single decision-maker~\cite{gawlikowski2023survey}

Our approach introduces sub-models to address these limitations. These sub-models are generated by selectively extracting representations from intermediate layers of the original model and introducing architectural diversity inspired by sub-ensemble methods.~\cite{gawlikowski2023survey} Each sub-model independently processes the same input, providing a detailed view of how the original model transforms inputs across layers. The key insight is that the degree of agreement among sub-model predictions correlates strongly with the trustworthiness of the model’s overall prediction. When sub-models produce consistent predictions, it indicates that the input is in-scope. Conversely, significant disagreement among sub-models suggests the input is likely out-of-scope.

%The key insight is the degree of agreement among sub-model predictions strongly correlates with the trustworthiness of the model’s overall prediction. Sub-models are generated by selectively extracting layer representations and introducing architectural diversity inspired by sub-ensemble methods~\cite{gawlikowski2023survey}. Each sub-model independently processes the same input, revealing how the original model transforms the input across layers. The more sub-models align with the main model's prediction, the higher the confidence that the prediction is accurate. Conversely, significant disagreement among sub-models suggests the input is likely out-of-scope.

Training sub-models focuses on capturing layer-specific uncertainties by freezing earlier layers and retraining only the dense layers, which act as classifiers for each task. This approach enhances the signal strength for determining model confidence. Additionally, studies have shown~\cite{maunveiling} that shallower layers in code models often outperform deeper ones for specific tasks, making the sub-model strategy particularly effective for code input validation. By dissecting predictions layer by layer, sub-models can pinpoint nuances in data representation, resulting in more accurate identification of in-scope and out-of-scope inputs.

\textbf{Sub-Model Generation.} 
Figure~\ref{fig:Sub_model} shows the overview of our Dropout-based Sub-Model Generation (DSMG). DSMG constructs diverse sub-models to capture the hidden state representations of the original model, enabling a more detailed analysis of input processing. Each sub-model consists two components: the first part is inherited from the original DL model, containing all structures and parameter values from the first layer up to an intermediate layer $k$ (for Sub-Model$_k$), and the second part is a newly trained dense layer linking layer $k$ to the output, customized for the specific software engineering (SE) task. 

 To generate first part, dropout-based hidden representations are extracted from each layer, introducing controlled randomness to highlight distinct processing behaviors. To generate second part, the dense layer is trained independently for each sub-model using cross-entropy as the loss function, which has demonstrated effectiveness for classification tasks in deep neural networks ~\cite{huang2017densely,he2016deep}. Dropout regularization is applied during training to prevent overfitting and enhance generalization. Notably, this sub-model generation process is performed offline, tailored to the SE task at hand, and does not affect the model during deployment.

%icse major revision Figure~\ref{fig:Sub_model} shows the overview of our Dropout-based Sub-Model Generation (DSMG). DSMG creates diverse sub-models to represent the hidden states of the original model.Each sub-model consists of two parts: the first part is derived from the original DL model, inheriting all structures and parameter values from the first layer up to an intermediate layer k (for Sub-Model$_k$), and the second part is a newly trained link from layer $k$ to the output layer, using training samples specific to each software engineering (SE) task.
%one part is obtained from the original DL model (i.e., for Sub-Model$_k$ is the first layer to intermediate layer k) by inheriting all structures and parameter values, \textcolor{blue}{and} the other part is the link from layer $k$ to the output layer which is trained using training samples \textcolor{blue}{of each SE task in our subject}.
 %icse major revision To generate first part, we obtained dropout-based layerwise hidden representation (illustrated below) of each layer. To generate second part, we adopted a dense layer, which has been proven to be effective in DNN for final layer prediction~\cite{huang2017densely,he2016deep} with labels from the original DL models. These sub models are trained using cross-entropy as its loss function. To mitigate overfitting and enhance generalization, we incorporated dropout regularization technique. Note that the sub-model generation is conducted only once in an offline way and are specific to each SE task.

\textit{Dropout-based layerwise hidden representation.}
In DSMG, hidden state representations are generated by selectively choosing specific layers and applying dropout to nodes within these selected layers. Dropout randomly omits a subset of nodes in a chosen layer during sub-model generation, while retaining the remaining active nodes. This approach introduces diversity along two axes: the depth of layers considered and the variation induced by dropout. Such diversity enables DSMG to effectively capture the model's processing of different input patterns and structures, providing a richer understanding of its behavior. This enhanced representation equips CodeImprove to better identify and differentiate in-scope inputs from out-of-scope inputs.

\begin{figure}[!htbp]
\centering
\includegraphics[width=\columnwidth]{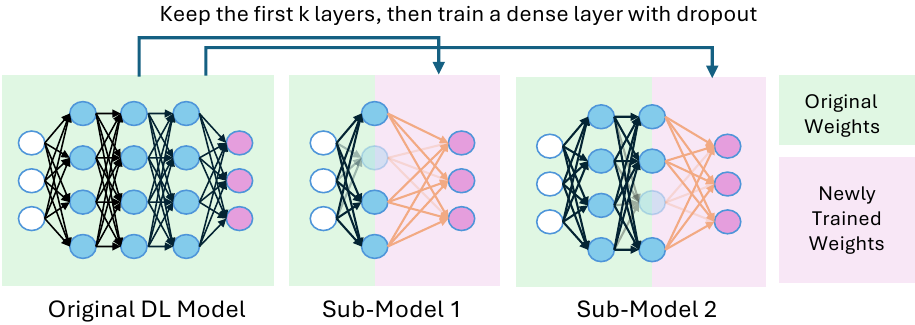}
\caption{Overview of Sub-Model Generation} 
% %\textcolor{red}{TODO: DRAW A FLOW DIAGRAM TO SHOW MANIFESTATION }
\label{fig:Sub_model}
\end{figure}

\textbf{Sub-model Validity Measurement.}
 Equation~\ref{eq:ic} outlines the computation of the validity score, which relies on understanding the processing of inputs across layers. For a given input $X$ fed into a DL model $M$ with $n$ labels, $M$ predicts $X$ as label $l_x$. Let $submodel_k$ be the softmax probability vector associated with the sub-model $k$. 

 To calculate the validity score, we differentiate between two scenarios: $l_x$ being a correct prediction or an incorrect one. For correct predictions, we employ the Best Versus Second Best (BVSB) strategy, which measures the difference between the highest predicted probability (labeled as $l_x$) and the second-highest predicted probability (labeled as $l_s$). This difference reflects the confidence of the sub-model, with a larger difference indicating higher confidence. For incorrect predictions, the BVSB strategy instead compares the actual highest predicted probability (labeled as $l_h$) with the probability assigned to the original model’s predicted label ($l_x$) highlighting the uncertainty between the sub-model’s confidence and the incorrect prediction. This approach, formulated in Equation~\ref{eq:bvsb} ensures that the validity score effectively captures the reliability of the sub-model’s predictions.

Once each sub-model computes the respective validity score for each input, we utilize the dissector's approach~\cite{wang2020dissector}  (i.e., weight growth types) to compute the validity for the whole DL model (Equation \ref{eq:pvscore}).

{\small
\begin{gather}
\label{eq:ic}
\text{ValidityScore}_k(l_x, \text{submodel}_k) = \\
\resizebox{0.98\columnwidth}{!}{
$
    \begin{cases}
       \text{submodel}_k[l_x] + B(\text{submodel}_k), & %\text{if } 
        l_x  \text{ with the highest probability} \\[5pt]%\text{submodel}_k  \\[5pt]
        \text{submodel}_k[l_x] -B(\text{submodel}_k), & \text{otherwise}
    \end{cases}
$
} \nonumber
\end{gather}
}

{\small
\begin{gather} 
\label{eq:bvsb}
    B(\text{submodel}_k) = \\
    \resizebox{0.8\columnwidth}{!}{
    $
    \begin{cases} 
      \text{submodel}_k[l_x] - \text{submodel}_k[l_s], & \text{if } l_x \text{ is correct}, \\
      \text{submodel}_k[l_h] - \text{submodel}_k[l_x], & \text{if } l_x \text{ is incorrect}.
    \end{cases} 
    $
    }\nonumber
\end{gather}
}

% {\small
% \begin{equation} \label{eq:bvsb}
%     B(\text{submodel}_k) = \text{submodel}_k[l_h] - \text{submodel}_k[l_s]
% \end{equation}
% }

{\small

\begin{equation}
\label{eq:pvscore}
\resizebox{\columnwidth}{!}{$
    Final_{score}(l_x) = \frac{\sum_{k=1}^{n} {ValidityScore_k}(l_x, submodel_k) \times weight_{k}}
{\sum_{k=1}^{n} weight_{k}}
$}
\end{equation}

}

\subsection{Input Adaptation} 

%\textcolor{red}{rewrite}
%\textcolor{red}{Add explanation about the transformation rules. }

Given an out-of-scope input, the goal of this phase is to covert the input to become an in-scope input. The challenge lies in exploring the large search space of possible program modifications while preserving semantic correctness. To tackle this, we define a set of semantic-preserving transformations. Guided by the DSMG validation score (Section~\ref{validation}), transformations are iteratively combined and refined to align inputs with the model’s capabilities. This process, formalized as Adaptation by Evolutionary Search (AES), efficiently optimizes the transformations to achieve the best solution.

%icse major revision Given an out-of-scope input, the goal of this phase is to covert the input to become an in-scope input. The challenge lies in identifying appropriate program transformations within the vast search space of possible modifications, which includes changes to code structure, logic, and data flow.  To address this challenge, we develop a list of basic semantic preserving transformations and leverage DSMG’s validation score as a guiding metric (Section~\ref{validation}) to combine these basic transformations into a composite transformation that effectively coverts the the input from being out-of-scope to in-scope. This method emphasizes iterative refinement guided by the metric, ultimately aiming to find the best solution. We named our approach Adaptation by Evolutionary Search (AES). 

 {
\begin{algorithm}
\caption{High-Level AES Algorithm}
\label{alg:GA}
\small
\DontPrintSemicolon
\SetKwInput{KwInput}{Input}
\SetKwInput{KwOutput}{Output}

\KwInput{Pre-trained Model $M$, Testing Dataset $T$, Out-of-scope Input ids $ids$, Maximum Iterations $max\_iter$, Mutation Rate $rate$, Fitness Threshold $threshold$}
\KwOutput{New Test program dataset $N$}

$N \leftarrow []$\;

\ForEach{$sample$ in $T$}{
    \If{$sample.id$ is not in $ids$}{
        $N.\text{append}(sample.code)$\;
        \textbf{continue}

    }

    $i \leftarrow 0$\;
    $best\_candidate \leftarrow sample.code$\;
    $fitness \leftarrow [0,\ldots,0]$ 

    \While{$i$ \textless{} $max\_iter$ \textbf{and} $fitness$[$best\_candidate$] \textless{} $threshold$}{
        $initial\_pop \leftarrow \text{genpop}(best\_candidate)$
        
        $fitness \leftarrow \text{fitness}(initial\_pop)$ 
        
        $new\_pop \leftarrow \text{select}(initial\_pop)$ 
        
        $pop \leftarrow \text{evolve}(new\_pop, rate)$

        $fitness \leftarrow \text{fitness}(pop)$\;
        
        $best\_candidate \leftarrow \text{select\_best}(pop)$ 
        
        $i \leftarrow i+1$\;
    }

    \If{$fitness$[$best\_candidate$] \textgreater{}$threshold$}{
        $N.\text{append}(best\_candidate)$
    }
    \Else{
        $N.\text{append}(sample.code)$
    }
}

\Return{$N$}\;
\end{algorithm}
}

\begin{table*}[h!]
\caption{List of Code Transformation}
\label{table:transformations}
\centering
\resizebox{0.9\linewidth}{!}{
\begin{tabular}{ c l l}
\hline
\textbf{No} &\textbf{Transformation Operator}  & \textbf{Description} \\
\hline
1 & changeName  & Function name and variable name renaming\\
2 & changeFor  &  The for-loop is transformed into a while-loop. \\
3& changeWhile & The while-loop is transformed into a for-loop \\
4 &changeDo  & The do-loop is transformed into a while-loop. \\
5 &changeIfElseIf  &Transformation of if elseif to if else \\
6&changeIf  &  Transformation of if else to if elseif\\
7& changeSwitch  & Transformation of the Switch statement to the if elseif statement. \\
8 & changeRelation & Transformation of relational expressions (e.g., a \textless{} b to b\textgreater{}a). \\
9& changeUnary & Modifications to unary operations (e.g., i++ to i = i+1) \\
10 & changeIncrement & Modifications to incremental operations (e.g., i+=1 to i = i+1). \\
11 & changeConstant & Modifying Constant (e.g., 0 \texttt{->} 8-8)  \\
12 & changeDefine & Modifications to variable definitions (e.g., int b=0 to int b; b=0).  \\
13 & changeAddJunk & Insert Junk Code that will never be executed. (e.g., if (0)\{printf();\})\\
14 & changeExchangeCod &\makecell[l]{Exchange the order of statements without dependencies (e.g., declaration statements)} \\
15 & changeDeleteComments &Deleting statements that print debugging
hints and comment.(e.g., printf()) \\
\hline
\end{tabular}}
\end{table*}

\textbf{Program Transformations.} To construct the set of basic program transformations, we considered all common kinds of code structures, including loop structures, branch structures, operator, and expression changes (Table ~\ref{table:transformations}). 
All transformations are carefully compiled to ensure that the adapted program not only undergoes transformation but also upholds the semantics of the original code. Our list of transformation operators is designed to be task-preserving by ensuring that the functionality and behavior of the code remain unchanged. %maintaining the semantic equivalence of the code, ensuring that the functionality and behavior of the code remain unchanged.} 
While there is potential to broaden this list of transformations, CodeImprove already shows significant improvement over existing techniques based on the current list. Due to the
space limitation, we list all these specific atomic operations at our project homepage~\cite{CodeImprove}. Then, we illustrate how to apply these operators to search for the best solution that deep code models can adapt.
\textbf{Adaptation by Evolutionary Search:}
One of the primary requirements before applying transformation operators is to identify syntactic features, i.e., the places of code fragments applicable for transformation. Finding appropriate syntactic features is essentially an optimization problem. CodeImprove addresses this problem by counting the number of code fragments present in each operator for a given code snippet. For example, if there are four identifiers in a source code snippet, then the count of the operator 1 is $K = 4$. 

After identifying the number of syntactic features to be transformed, CodeImprove needs a transformation strategy to generate a diverse pool of candidates. CodeImprove achieves this by implementing a genetic algorithm-based~\cite{kumar2010genetic,forrest1996genetic,whitley1994genetic} strategy comprising initialization, a fitness function, crossover and mutation operators, and termination criteria to guide the transformation process.  Algorithm~\ref{alg:GA} shows the overview of our transformation strategy. 
The inputs for the AES algorithm are the trained DL model $M$, the test data used to evaluate the $M$'s performance, the identified out-of-scope data, the maximum number of iterations that the AES should evolve for, and the required fitness score that the solution of AES should achieve. The output of our genetic algorithm is a new dataset that includes the transformed source code. To avoid randomness when applying transformations, CodeImprove transforms all $K$ counts of features in each operator. 
%The output of our genetics algorithm is to create a new  dataset that includes the transformed source code. Note that to avoid randomness when applying transformations, CodeImprove transforms all $K$ counts of features in each operator. 

Initially, CodeImprove creates a starting population by applying 15 operators to each out-of-scope input (Line 10), with each individual in this population representing a potential solution. The fitness of each individual, determined by the DSMG's validation score (Section~\ref{validation}), reflects how closely a solution approaches the problem's target, with higher scores indicating better candidates (Line 11).

% \textcolor{red}{Applying traditional genetic algorithm for code data is challenging. First, genetic algorithm for code transformation should prioritize semantic preservation. Traditional approaches may operate on string binary, binary or numerical representations~\cite{kumar2010genetic,forrest1996genetic,whitley1994genetic} cannot be directly used as for code data, the focus is the specific code syntax that needs to be transformed. Hence the representation of code data should either be the code snippet itself or the abstract syntax trees (AST). Second, the fitness functions in genetics algorithms are are simple objective functions~\cite{forrest1996genetic}, however, for code data, it is required a metric to guide the transformation to enhance the model's performance. Third, traditional approaches, crossover techniques are focused on single-point, two-point, or uniform crossover~\cite{kumar2010genetic,forrest1996genetic,whitley1994genetic}. However, the crossover operator for code data should be the set of transformations applied to the code snippet and operate directly on the code structure, enabling transformations at statement, expression, or program construct level. Such variations poses challenges in leveraging genetics algorithms for code data. }

Then, the genetic algorithm iterates through cycles of evaluation, selection, and reproduction.  In the selection phase, the top 50\% of candidates are chosen based on their fitness scores (Line 12). During the reproduction phase, CodeImprove performs genetic operators (i.e., crossover and mutation) to generate new
solutions (Line 13). During crossover, for each candidate, CodeImprove applies a sequence of transformations. %(i.e., each candidate in the population will undergo 1, 5, 10, and 15 transformations). 
Then, we add the new samples to our population. For each crossover variant, we mutate the syntactic feature with a random transformation to maintain a diverse population. The algorithm terminates when the candidate code has reached a higher validation score based on our guiding metric or the fixed number of generations has reached. In the end, the algorithm returns the solution with the highest
fitness value (Line 15). %After obtaining the new testing dataset, we input the dataset to identify the performance of program transformation of CodeImprove.
Next, we will describe the experimental evaluation.

\section{Evaluation}
\label{evaluation}
\begin{table*}[t!]
\caption{Effectiveness of CodeImprove}
\label{Tab:performance}
\resizebox{\linewidth}{!}{
\begin{tabular}{cccccccccccccc}
\hline
 \multirow{2}{*}{\textbf{Experiment}}  & \multirow{2}{*}{\textbf{Model}} &\multicolumn{5}{r}{\textbf{Vulnerability Detection}} & \multicolumn{5}{r}{\textbf{Defect Prediction}}\\ \cmidrule(lr){3-8} \cmidrule(lr){9-14} %\cline{3-12}
    &  &A(\%) & P(\%) & R(\%) & F1(\%) &RI(\%)  &CSR(\%$\uparrow$)/MCR(\%$\downarrow$)& A(\%) & P(\%) & R(\%) & F1(\%) &RI(\%) &CSR(\%$\uparrow$)/MCR(\%$\downarrow$)\\ \hline

    \multirow{3}{*}{Original set up} & CodeBERT & 62.74 & 62.31  & 47.81  & 54.11 &- & - &81.98 & 82.12 & 81.98& 81.45 & - & - \\

    & RoBERTa & 61.56 & 57.71 & 61.11 & 59.36 & - & - & 80.02 & 79.91 & 80.02 & 79.40&-&- \\

    & GraphCodeBERT & 62.40 & 61.50 & 48.20 & 54.09 &- & -&81.91 &  81.77& 81.91& 81.56&- & -\\\hline

    \multirow{3}{*}{ InputReflector} & CodeBERT & 62.48  &61.97& 47.4  & 53.72 & -1.4 & 1.6/0.05 &80.85  &81.01 &80.85 & 80.24 &  -19.4 &2.4/5.4 \\

    & RoBERTa &62.48  &60.6  & 52.2 &56.1  & 9.4&21.3/6.59& 79.73 &79.59 &79.73 & 79.03 &-6.6 & 0.3/0.4   \\

    & GraphCodeBERT&62.81 &60.34  & 55.53 & 57.84 & 4.9 &18.0/4.7 & 80.32&80.13 &80.32&79.92 &-21.8  &1.42/2.1  \\\hline

    % \multirow{3}{*}{ IT} & CodeBERT &  &  && &  &  &  & & &  &  &  \\\cline

    %& RoBERTa &  &  & & & &&  &  & & &  &   \\\cline

   % & GraphCodeBERT& &  &  &  & & & && &  &  &  \\\hline

    \multirow{3}{*}{ CodeImprove} & CodeBERT & \textbf{71.52} 
      & \textbf{70.79}  & \textbf{64.70}&\textbf{67.61}&\textbf{51.28} & \textbf{39.9/2.6} &\textbf{84.53}  &\textbf{84.22} &\textbf{84.43}&\textbf{83.99}  & \textbf{44.2} &  \textbf{32.8/1.0}\\

    & RoBERTa &\textbf{68.81} &\textbf{64.67}  &\textbf{70.76}  &\textbf{67.58}  &\textbf{74.4} &\textbf{36.2/1.96} &\textbf{82.91} &\textbf{82.99}&\textbf{82.91} &\textbf{82.36} &\textbf{65.6}  & \textbf{32.2/1.0}  \\

    & GraphCodeBERT &\textbf{65.26}  & \textbf{64.22} &\textbf{55.06} &\textbf{59.29} &\textbf{35.1}  &\textbf{23.1/1.9} &\textbf{83.45} &\textbf{83.55} & \textbf{83.45}& \textbf{83.09} &\textbf{21.1}  & \textbf{27.2/1.4} \\\hline

\end{tabular}}
\end{table*}

We conducted all our experiments on a server equipped with an Intel Xeon E5-26 CPU and eight NVIDIA 1080 Ti GPUs. We set up 0.3 and 0.2 as the threshold based on our validation score on detecting out-of-scope inputs for vulnerability detection and defect prediction tasks, respectively, on all subjects. We set up maximum iteration to three and different crossover rates (i.e., 0.16, 0.33, 0.66, and 1)  when applying sequences of transformation operators on each subject matter described in Section~\ref{evaluation}.

%We set up 0.3 and 0.2 as the threshold based on our validity score on detecting out-of-scope inputs for vulnerability detection and defect prediction tasks respectively on all subjects. We set up maximum iteration to three, and crossover rate of 0.5 when applying program transformations on each subject matter. 

\subsection{Research Questions }
%We aim to answer the following research questions in the next section:

\noindent\textbf{RQ1. Overall Performance: }What is the overall performance of CodeImprove?

\noindent\textbf{RQ2. Input Validation: } How effective is the out-of-scope program data detection? 

\noindent\textbf{RQ3. Input Adaptation: } How effective to convert out-of-scope data to become in-scope data?

\noindent\textbf{RQ4. Setting Sensitivity: }
How sensitive is CodeImprove's performance under different experimental setups? 

%%\noindent\textcolor{blue}{\textbf{RQ5. Submodel Decomposition: }How effective is the submodel generation in detecting out-of-scope data detection?}

%\textbf{RQ3. Layer Selection:} Does the choice of layers for selection have an impact on out-of-scope data detection?
\noindent\textbf{RQ5. Semantic Preserving:} Does program transformations of CodeImprove preserve semantics?

%\textbf{RQ5. Transformation:} How effective are the program transformations on out-of-scope data?

\noindent\textbf{RQ6. Runtime Overhead:} What is the overhead of CodeImprove in adapting a program to DL models?

%\textbf{RQ6. Controlling Factors: } What factors contribute to the impact on the performance of DL models?

%\textbf{RQ2.} What is the accuracy of the model with transformed programs?

\subsection{Subjects}
\label{subjects}
\subsubsection{Datasets and Tasks}
CodeImprove is evaluated on two code-based classification tasks: vulnerability prediction with devign dataset ~\cite{zhou2019devign}, and defect prediction with codeChef dataset~\cite{phan2017conv}. Although this study used two datasets, CodeImprove applies to all code model-based tasks, with plans for broader future evaluations.

%{While these two datasets were selected for this study, CodeImprove is applicable to all code model-based tasks, and we plan to evaluate CodeImprove on a broader range of tasks in the future.}

%To evaluate CodeImprove, we consider two code-based classification tasks and two associated datasets (i.e., vulnerability prediction with devign dataset ~\cite{zhou2019devign}, and defect prediction with codeChef dataset~\cite{phan2017conv}).   

%The statistics of datasets are shown at the first four columns in Table~\ref{table:datasets}, each of which represents the task, the number of classes for the classification task, the class names, the number of samples in each class, the number of inputs in the training/validation/test set, and the programming language for the inputs. 

For \textbf{vulnerability detection}, the Devign dataset\cite{zhou2019devign} consists of 27,318 functions extracted from FFmpeg and Qemu open-source C projects. The functions are labeled as containing vulnerabilities or being clean, with 14,858 vulnerable and 12,460 clean samples. The dataset is split into train/validation/test sets with sizes 21,854/2,732/2,732.

For \textbf{defect prediction}, the CodeChef dataset includes 33,822 C/C++ functions from the CodeChef platform~\cite{phan2017conv}. Samples are labeled with categories such as
%Samples are labeled as having 
no defect, wrong output, timeout error, or runtime error, with 11,362 no defect, 13,656 wrong output, 5,101 timeout error, and 3,703 runtime error samples. The dataset is divided into train/validation/test sets with sizes 21,647/5,411/6,764.

%The task of \textbf{vulnerability detection} aims to predict whether a given code snippet contains vulnerabilities. We use the dataset that was prepared by \cite{zhou2019devign}. The dataset is extracted from two popular open-sourced C projects: FFmpeg and Qemu, consisting of 27,318 functions that are labeled as either containing vulnerabilities or clean. The number of vulnerable samples are 14858 and clean samples are 12460. The dataset consists of 21854/2732/2732 as train/validation/test samples. The task of \textbf{defect prediction} aims to predict whether a given code snippet is defective and its defect type. We use the CodeChef dataset that was prepared by ~\cite{phan2017conv}. The dataset is extracted from the CodeChef platform that contains 33822 C/C++ functions that are labeled as no defect, wrong output, timeout, or runtime error. The dataset consists of 11362 no defect samples, 13656 wrong output samples, 5101 timeout error samples, and 3703 runtime error samples. The dataset consists of 21647/5411/6764 as train/valid/test samples. 

%This dataset is included as part of the CodeXGLUE benchmark~\cite{} that has been used to investigate the effectiveness of CodeBERT for vulnerability prediction. CodeXGLUE divides the dataset into training, development and test set that we reuse in this study. 

%\input{tex/tables/datasets}

\subsubsection{Models}
We employed state-of-the-art pre-trained models, namely CodeBERT~\cite{fengCodeBERTPreTrainedModel2020}, RoBERTa~\cite{Liu2019RoBERTa}, and GraphCodeBERT~\cite{guo2021graphcodebert} which have been widely utilized in previous studies~\cite{lu2021codexglue,tian2023code,yang2022natural,carrot}. These models were fine-tuned on our tasks using the corresponding datasets, adhering to the recommended settings proposed in previous literature~\cite{lu2021codexglue}. Hyper-parameters were set to match the original configurations. CodeImprove is designed for use with all types of code models. Our study includes a diverse range of tasks, pre-trained models, and class numbers, ensuring a comprehensive evaluation of CodeImprove's performance.

\subsection{Evaluation Metrics }

We use a diverse set of metrics to measure CodeImprove’s effectiveness for our six RQs.   

%\textbf{Accuracy:} Accuracy = $\frac{TP+TN}{TP+TN+FP+FN}$. 
The \textbf{accuracy (A)} is the proportion of correctly classified samples out of all samples. %TN represents the number of true negatives and TP + TN + FN + FP represents the number of all samples.
 %\textbf{Precision:} Precision = $\frac{TP}{TP+FP}$. 
The \textbf{precision (P)} is the percentage of correctly predicted positive samples out of all positive predictions. %TP and FP denote the number of true positives and false positives, respectively. 
%\textbf{Recall:} Recall = $\frac{TP}{TP+FN}$. 
The \textbf{recall (R)} measures the percentage of correctly predicted positive samples that were retrieved out of all actual positive samples.  %TP and FN denote the number of true positives and false negatives, respectively.
%\textbf{F1-Score:} F1 = $\frac{2*Precision*Recall}{Precision+Recall}$ = $\frac{2*TP}{2*TP+FP+FN}$. 
The \textbf{F1 score (F1)} is the harmonic mean of precision and recall. 
\textbf{Relative improvement (RI)} quantifies the accuracy improvement relative to the difference between training and test accuracy.

%is measured by the amount of accuracy improved over the difference between training accuracy and test accuracy.

% \textbf{Relative Improvement: }RI = $\frac{\text{New Accuracy} - \text{Orginal Accuracy}}{\text{Training accuracy- Original Accuracy}}*100$

\textbf{ Correction success rate (CSR)}~\cite{tian2023code}
is the ratio of successfully corrected mispredictions to the total identified mispredictions. \textbf{Mis-correction rate (MCR)}~\cite{tian2023code} measures the negative effect caused, which is the ratio of correct predictions changed to mispredictions to the total number of correct predictions in the test set.

%measures the ability to correct mispredictions, which is the ratio of the number of inputs whose mispredictions are successfully corrected to the total number of identified mispredicted inputs. \textbf{Mis-correction rate (MCR)} measures the negative effect caused, which is the ratio of the number of inputs whose correct predictions are changed to mispredictions to the total number of inputs with correct predictions in the test set.

\textbf{ Correction validation rate (CVR)} is the ratio of successfully validated mispredictions to the total possible mispredicted inputs to be validated. \textbf{Mis-correction validation rate (MVR)} is the ratio of correct predictions validated as mispredictions to the total number of correct predictions in the test set. \textbf{AUC} score evaluates the effectiveness of the input validation process. \textbf{Transformations per Second (TPS)} measures the rate of transformations CodeImprove can apply per second. Next, we will describe the results of the experiments.

\section{Results and Analysis}
\label{results}
We report and analyze experimental results, and answer the preceding research questions in turn.

\subsection{RQ1: Overall Performance of CodeImprove}
%\input{tex/tables/evaluation}

%We discuss the results of the effectiveness of applying program transformations on the out-of-scope data. Table ~\ref{Tab:evaluation} represents the statistics of our experimental results. In addition to Table~\ref{Tab:evaluation}, we also show the statistics of number of inputs that program transformation has been applied, the number of correctly predicted and incorrectly predicted inputs of the total inputs by comparing those to the ground truth data, and number of transformed inputs that convert a correct prediction to incorrect vice versa in Table~\ref{Tab:stts}.

\textit{Baseline}: CodeImprove is the first technique to improve the code model's performance through program transformations. Thus, we cannot find direct baselines for comparison. To address this, we draw inspiration from a technique in the image domain for comparative analysis, i.e. the InputReflector \textbf{(IRef)}~\cite{xiao2022repairing} that detects deviating inputs and substitutes them with the most similar sample from the training set. We are unable to include CodeDenoise~\cite{tian2023code} as a baseline due to the evaluation methods of this work, which splits the test set into two subsets as T1 and T2 and subsequently assesses results on T2 set. However, the splitting criteria of datasets is not provided in the project website. Moreover, the work is similar to the adversarial style, which denoises the program identifiers and corrects the program with the supervision of the model's predictions. CodeImprove demonstrates better performance compared to CodeDenoise~\cite{tian2023code} where it only fixes 20.45\% of inputs while CodeImprove fixes 32.8\% for defect prediction task on the CodeBERT model.

%We borrowed the ideas from two techniques in the image domain for comparison. Input reflector (IR)~\cite{xiao2022repairing} and interpretable technique (IT)~\cite{mohasel2023interpretable} identifies deviating inputs and replace with the samples in training set. %We ignore the work in ~\cite{tian2023fly} because the work identifies noisy identifiers in the dataset by applying perturbations and then clease these identifiers. Although, such technique can help in improving model performance, it requires to generate multiple samples for input validation by applying perturbations for multiple identifiers. Moreover, the work splits the test set into two sets, thus making the sample size even smaller. 

\textit{Process}: \textbf{IRef} utilizes two models, the siamese network~\cite{he2018twofold} and the quadruple network~\cite{xiao2022repairing}, to detect deviating inputs and repair them. During training, these models rely on three datasets: the original set, a transformed set (human recognizable), and an extremely transformed set (human unrecognizable). These transformed sets are created by applying varying degrees of transformation to the original data. However, for code data, generating such datasets is challenging as transformations do not follow a continuous degree like in image data. Therefore, we utilized only two sets: the original set and a transformed version.

We adapted the IRef loss functions for these two sets and fed hidden layer outputs from the original model into the siamese and quadruple networks. To repair out-of-scope inputs, IRef searches for the most similar data in the training set and exchanges their labels. In contrast, CodeImprove applies all 15 semantic-preserving transformations during the crossover step. Effectiveness is evaluated using metrics such as accuracy, precision, recall, F1-score, RI, CSR, and MCR.

%\textbf{IRef} employs two models, namely the siamese network~\cite{he2018twofold} and the quadruple network~\cite{xiao2022repairing}, to detect deviating inputs and repair them, respectively. During training, these models rely on three datasets: the original set, the transformed (human recognizable) set, and the extremely transformed (human unrecognizable) set. These two transformed sets are generated by applying different degrees of transformation to the original training set. However, generating such datasets for code data poses challenges as code inputs cannot rely on a degree of transformation like in image data. Therefore, we utilized only two sets: the original set and its transformed version. We configured the loss functions employed in the IRef accordingly for these two sets. We leverage the hidden layer outputs from the original model and feed them to the siamese network and quadruple network. Subsequently, we search for similar data in the training set to repair the out-of-scope inputs by exchanging the labels of the most similar training sample. For CodeImprove, we employed all 15 transformations during the crossover step. The effectiveness is measured using various metrics, including accuracy, precision, recall, F1-score, relative improvement (RI), correction success rate (CSR), and mis-correction rate (MCR).   
%%edited Table~\ref{Tab:performance} presents a detailed comparison between CodeImprove and \textcolor{blue}{IRef}, illustrating that CodeImprove consistently outperforms \textcolor{blue}{IRef} across various metrics. 

\textit{Result: }Table~\ref{Tab:performance} presents the comparison between CodeImprove and IRef, illustrating that CodeImprove consistently outperforms IRef. 
%results comparing CodeImprove with IR. From this table, we found that CodeImprove always performs better than IR.
Notably, we observe the following: (1) CodeImprove consistently achieved the best model improvements with upto 8.78\% in accuracy, 8.48\% in precision, 16.9\% in recall, and 13.5\% in F1-score on all the subjects; (2) CodeImprove is capable of correcting around 23.1\% to 39.9\% of the mispredicted inputs on both vulnerability detection and defect prediction tasks; (3) CodeImprove shows notable RI improvements, ranging from 21.1\% to 51.28\%, particularly excelling with RoBERTa models; 
%The RI measurement shows significant improvement for RoBERTa models on both the tasks. On other models, CodeImprove shows a RI ranging from 21.1\% to 51.28\%; 
(4) Techniques designed for image data (e.g., \textbf{IRef}) fail in the context of code. IRef negatively impacts performance, especially for CodeBERT, as it focuses on out-of-distribution inputs rather than inputs prone to misprediction within the same distribution. This limitation arises from IRef’s inability to effectively handle the syntactic and semantic similarities between transformed and original code; and 
%%Techniques employed in the image domain \textbf{(IRef)} cannot be applied to code data. Our results indicate that IRef negatively impacts the, 
%as our results show that IR hurts
 %%model performance for the CodeBERT model on both vulnerability detection and defect prediction tasks. Moreover, the defect prediction task shows performance discrepancy on all models for \textbf{IRef}. One of the reasons is that, \textbf{IRef}'s analyzer component is better at detecting %uncertainty measurements~\cite{hu2023codes,li2021estimating} are for detecting 
%%out-of-distribution (i.e., data from a different distribution from training data) inputs, and is able to repair such inputs. %extremely transformed images.
%%However, our aim is to maintain the same distribution while identifying inputs that are prone to misprediction. Thus, \textbf{IRef} fails to capture out-of-scope inputs adequately due to the similarity in syntax and semantics between transformed and original code; and
%So IR fails to capture out-of-scope inputs because a transformed version of a program can have similar syntax and semantics compared to original code; 
(5) CodeImprove introduces minimal negative effects, with only 2.6\% of correct predictions misclassified in the worst case. CodeImprove successfully adapts out-of-scope inputs to in-scope inputs, as demonstrated in Table~\ref{Tab:performance}.
\begin{tcolorbox}[title=\textbf{RQ1} - What is the overall performance of CodeImprove?, left=2pt, right=2pt,top=2pt,bottom=2pt]
CodeImprove was effective in adapting out-of-scope inputs for both SE tasks on three subject models with higher accuracy, precision, recall, F1-score, CSR, and RI.(Table~\ref{Tab:performance}). 
\end{tcolorbox}

\subsection{RQ2: Effectiveness of Out-of-Scope Data Detection}
\label{RQ2}

%\textcolor{red}{Add figures for uncertainty scores, and descriptions}
%%%We evaluate the effectiveness of out-of-scope data detection using the AUC value. Specifically, we compute the AUC values of three variants of dissector that are designed according to different weight growth types (i.e., Dissector-linear $(y = x)$, Dissector-log $(y = ln x)$, and Dissector-exp $(y = e^x )$ and apply the same three variants on our validity score analysis. Table~\ref{Tab:rq2} shows the statistics of the AUC comparison of the three variants of the Validity Score with the Dissector. 
\textit{Baseline:} We compared CodeImprove's DSMG with the Dissector~\cite{wang2020dissector}. 
Additionally, we evaluated the DSMG approach with the uncertainty metrics in our preliminary study (Section ~\ref{study}): Vanilla, temperature-scaling, predictive entropy, entropy, mutual information, least confidence, ratio confidence, and margin confidence, monte-carlo dropout, and deep ensemble. %\textcolor{blue}{For} a more comprehensive study on other uncertainty metrics, \textcolor{blue}{please refer to our project website}~\cite{Data}. %is added to our project website~\cite{Data}. 

\textit{Process:} We compute the AUC score, CVR, and MVR to evaluate the effectiveness of out-of-scope data detection on all baseline approaches.

\begin{table}[htb!]
\caption{Effectiveness of Input validation}
\label{Tab:validation}
\renewcommand{\arraystretch}{}
\resizebox{\linewidth}{!}{
\begin{tabular}{ccccccccc ccccc}
\hline
 \multirow{2}{*}{\textbf{Experiment}}  & \multirow{2}{*}{\textbf{Model}} &\multicolumn{2}{c}{\textbf{Vulnerability Detection}} & \multicolumn{2}{c}{\textbf{Defect Prediction}}\\ \cmidrule(lr){3-4} \cmidrule(lr){5-6} % {3-12}
    &  &CVR(\%$\uparrow$)/MVR(\%$\downarrow$) & AUC &  CVR(\%$\uparrow$)/MVR(\%$\downarrow$) & AUC \\ \hline

    \multirow{3}{*}{Vanilla}  &CodeBERT &34.7/32.1  &0.552 &32.4/22.2 &0.595 \\

    & RoBERTa  &13.7/21.2 &0.574 &10.6/8.7 & 0.580   \\

    & GraphCodeBERT  &20.6/16.9 & 0.455 &18.8/20.8 & 0.494   \\\hline

    \multirow{3}{*}{ Temp. Scaling} & CodeBERT  & 38.8/31.8&0.420 &36.5/27.6 & 0.403\\

    & RoBERTa  &10.0/15.4 &0.572 &28.5/16.2 &  0.413  \\

    & GraphCodeBERT& 15.8/11.9 & 0.442& 18.2/19.3& 0.493 \\\hline

     \multirow{3}{*}{ Predictive Entropy} & CodeBERT  & 43.5/35.3&0.582 &12.3/4.6 & 0.573 \\

    & RoBERTa & 28.3/23.1& 0.454&42.8/30.1 & 0.579   \\

    & GraphCodeBERT& 13.1/14.6 & 0.508 &25.6/29.0 &0.491 \\\hline

    \multirow{3}{*}{ Entropy} & CodeBERT  &12.5/30.8 & 0.414& 39.1/27.8&  0.436\\

    & RoBERTa &7.1/8.6 &0.556 & 29.9/16.7&  0.391  \\

    & GraphCodeBERT & 16.1/17.7&0.482 &23.2/16.4 &0.426  \\\hline

    \multirow{3}{*}{ Mutual Information} & CodeBERT  & 8.6/25.7&0.586 & 30.7/22.2& 0.566 \\

    & RoBERTa & 25.7/21.1&0.443 &44.4/30.1 &  0.608  \\

    & GraphCodeBERT &35.2/41.1 &0.531 & 19.2/14.5&0.579  \\\hline

    \multirow{3}{*}{ Least Confidence} & CodeBERT  & 16.7/14.5&0.589 &19.1/9.7 &0.595  \\

    & RoBERTa & 24.7/32.1& 0.452& 10.7/13.9&  0.593  \\

    & GraphCodeBERT &22.0/18.0 &0.558 &16.7/20.2 & 0.508 \\\hline

    \multirow{3}{*}{Ratio Confidence} & CodeBERT  & 8.4/14.5&0.464 & 11.7/11.5&0.553  \\

    & RoBERTa &17.1/14.4 & 0.521& 6.3/9.7&  0.484  \\

    & GraphCodeBERT &47.3/42.8 &0.486 &26.9/21.7 & 0.469 \\\hline

    \multirow{3}{*}{ Margin Confidence} & CodeBERT  & 6.6/11.2&0.473 &10.6/10.1 & 0.562 \\ 

    & RoBERTa & 37.2/28.5& 0.521&9.2/5.7 &   0.532 \\ 

    & GraphCodeBERT &35.1/32.6 & 0.489& 8.3/5.1&0.470  \\\hline

   \multirow{3}{*}{ MCD} & CodeBERT  & 28.3/16.8 &0.624 &17.9/4.9 & 0.614  \\ 
   
   & RoBERTa &36.2/34.1 &0.613 & 28.2/7.3&   0.607 \\ 
    
    & GraphCodeBERT &38.7/30.3 & 0.617 & 20.5/5.3 &0.616  \\\hline

    \multirow{3}{*}{ DE} & CodeBERT  & 44.4/43.9 &0.507 &24.7/20.8 & 0.561  \\ 
   
   & RoBERTa &48.1/33.5 & 0.519 & 24.2/17.7&   0.562 \\ 
    
    & GraphCodeBERT &36.1/38.1 & 0.507& 23.2/19.1 &0.571  \\\hline

    \multirow{3}{*}{ Dissector} & CodeBERT  & 68.7/15.1& 0.850& 53.3/3.6& 0.889 \\ 

    & RoBERTa &54.3/14.8 &0.819 &47.1/5.4 &  0.828  \\ 

    & GraphCodeBERT &40.1/16.7 &0.757 &53.4/4.6 & 0.873 \\\hline

    \multirow{3}{*}{ CodeImprove} & CodeBERT  &\textbf{70.4/13.7} & \textbf{0.876}&\textbf{57.9/3.0} &\textbf{0.911}  \\ 

    & RoBERTa &\textbf{60.4/14.7} &\textbf{0.825} & \textbf{58.1/3.1}&  \textbf{0.924}  \\ 

    & GraphCodeBERT &\textbf{47.2/13.0} &\textbf{0.781} &\textbf{56.1/3.3} &\textbf{0.909}  \\\hline
 % MCD & 0.624 &0.613 & 0.617 & 0.614 & 0.607 &0.616 \\\hline

    %DE  & 0.507 & 0.519 & 0.507 & 0.561 & 0.562 & 0.571 \\\hline
  
\end{tabular}}
\end{table}

\begin{table}[htb!]
\caption{Effectiveness of Sub-model Decomposition}
\label{Tab:hiddenvalidation}
\renewcommand{\arraystretch}{}
\resizebox{\linewidth}{!}{
%\begin{tabular}{cccc}
\begin{tabular}{>{\color{black}}c>{\color{black}}c>{\color{black}}c>{\color{black}}c}
\hline
 %\multirow{2}{*}{\textbf{Experiment}}  & \multirow{2}{*}{\textbf{Model}} &\multicolumn{1}{c}{\textbf{Vulnerability Detection}} & \multicolumn{1}{c}{\textbf{Defect Prediction}}\\ \cmidrule(lr){3-3} \cmidrule(lr){4-4} %  {3-12} &   & AUC &  AUC \\ \hline

 \multirow{2}{*}{\textbf{Experiment}}  & \multirow{2}{*}{\textbf{Model}} &\multicolumn{1}{>{\color{black}}c}{\textbf{Vulnerability Detection}} & \multicolumn{1}{>{\color{black}}c}{\textbf{Defect Prediction}}\\   {3-4} 
    &   & \textbf{AUC} &  \textbf{AUC} \\ \hline

    \multirow{3}{*}{Hidden States}  &CodeBERT &0.557 &0.607 \\  

    & RoBERTa  &0.534 & 0.582   \\  

    & GraphCodeBERT  & 0.451 & 0.593   \\\hline

    \multirow{3}{*}{CodeImprove} & CodeBERT  & \textbf{0.876} & \textbf{0.911}\\  

    & RoBERTa  &\textbf{0.825} & \textbf{0.924}  \\  

    & GraphCodeBERT&  \textbf{0.781}&  \textbf{0.909} \\\hline

\end{tabular}}
\end{table}%on different uncertainty methods compared to with DSMG approach. %Specifically, we compute the AUC values of three variants of dissector~\cite{wang2020dissector} that were designed according to different weight growth types (i.e., Dissector-linear $(y = x)$, Dissector-log $(y = ln x)$, and Dissector-exp $(y = e^x )$. We applied the same three variants on our validity score analysis. The statistics of the AUC comparison of the three variants of the validity score with the dissector are presented in Table ~\ref{Tab:rq2}.
%\vspace{-1em}
\textit{Results: } Table ~\ref{Tab:validation} shows the results of out-of-scope data detection. Based on the results, we observe that: (1) CodeImprove achieved higher AUC scores across all models and tasks (i.e., AUC 0.781- 0.924); (2) While Dissector performs better than other uncertainty metrics, CodeImprove still surpasses it in AUC scores; 
%Although dissector obtained better AUC scores compared to other uncertainty metrics, CodeImprove still outperforms dissector; 
(3) CodeImprove demonstrates a higher CVR across all subjects, detecting up to 70.4\% of out-of-scope inputs for CodeBERT on vulnerability detection tasks and consistently outperforming other methods in CVR;  
%The CVR on DSMG is higher than all other baselines. Moreover, CodeImprove can detect 70.4\% of the out-of-scope inputs for the CodeBERT model on vulnerability detection tasks. CodeImprove consistently outperforms other techniques in terms of CVR on each subject.; 
(4) The MVR on CodeImprove is lower than other approaches, concluding that CodeImprove is better at differentiating in-scope inputs. MVR for defect prediction task shows 3.0\%, 3.1\%, and 3.3\% for CodeBERT, RoBERTa, and GraphCodeBERT models; (5) MCD and DE average predictions over multiple forward passes through the same network. This approach limits diversity in the predictions, which may contribute to their poorer performance in AUC compared to CodeImprove; and (6) other uncertainty metrics did not produce promising results on AUC, CVR, or MVR. For example, predictive entropy obtained a CVR of 43.5\% and an MVR of 35.3\%, which are not significant indicators of effective performance.

\textit{\textbf{Comparison of Using  Sub-models vs. Hidden State Outputs from the Original Model:}}
%\textcolor{blue}{Although the submodels are trained by leveraging hidden representation of the original model, accessing the hidden representation directly is possible. Therefore, we conduct an experiment to show the effectiveness of using trained sub-models compared to direct use of hidden representations for out-of-scope data detection.} 
%\textcolor{red}{make it quite clear sub-model decomposition vs hidden states of original model. add epistemic: dropout and emsemble. }
Sub-models are trained using the layerwise hidden states of the original model, but it is also possible to directly access these hidden states for detecting out-of-scope inputs. We conducted an experiment comparing trained sub-models with the direct use of hidden states. Due to the high dimensionality of the hidden states, we applied a linear transformation and subsequently applied Equations ~\ref{eq:ic} - ~\ref{eq:pvscore}. Table ~\ref{Tab:hiddenvalidation} presents the statistics of the AUC comparison between these two methods.

Based on the results in  Table~\ref{Tab:hiddenvalidation}, we observe that training sub-models (AUC 0.781-0.924) outperforms direct use of hidden states (AUC 0.451 - 0.607) across both software engineering tasks. These results signify the performance of trained sub-models for out-of-scope input validation. We summarize several factors contribute to the lower effectiveness of directly using hidden states: 1) \textbf{Ineffective Feature Utilization and Transformation}: %Hidden states are in the form of \textit{(batch\_size, sequence\_length, hidden\_size)}. 
Effective input validation requires a mapping between the feature space and the class space. Without training a dense layer, this process would merely reduce dimensions without learning this mapping. Training a dense layer allows it to learn the most relevant features from the hidden states and establish an accurate mapping from the feature space to the class space. This reduces significant information loss and enhances overall performance; and 2) \textbf{Lack of Adaptability}: Trained dense layer in sub-models can adapt to the characteristics and distribution of the training data, making them more effective for each SE task. Without training, the model lacks this adaptability, resulting in poor effectiveness.

\begin{tcolorbox}[title=\textbf{RQ2} - How effective is the out-of-scope program data detection?, left=2pt, right=2pt,top=2pt,bottom=2pt]
CodeImprove effectively distinguishes out-of-scope from in-scope inputs (AUC: 0.781-0.924, CVR: 47.2\%-70.4\%, MVR: 3.0\%-14.7\%), and is more suitable than existing techniques for various SE tasks.
%CodeImprove was effective in distinguishing out-of-scope inputs from in-scope inputs (AUC: 0.781-0.924, CVR: 47.2\%-70.4\%, and MVR: 3.0\%-14.7\%), and more suitable than existing techniques for different SE tasks with varying accuracies.
\end{tcolorbox}
\vspace{-1em}

\begin{table*}[htb!]
\caption{Effectiveness of Search Strategy for Input Adaptation}
\label{Tab:evaluation}
\renewcommand{\arraystretch}{}
\resizebox{\linewidth}{!}{
\begin{tabular}{ccccccccc ccccc}
\hline
 \multirow{2}{*}{\textbf{Experiment}}  & \multirow{2}{*}{\textbf{Model}} &\multicolumn{5}{r}{\textbf{Vulnerability Detection}} & \multicolumn{5}{r}{\textbf{Defect Prediction}}\\ \cmidrule(lr){3-8} \cmidrule(lr){9-14} % {3-12}
    &  &A(\%) & P(\%) & R(\%) & F1(\%) &RI(\%$\uparrow$)  &CSR(\%$\uparrow$)/MCR(\%$\downarrow$)& A(\%) & P(\%) & R(\%) & F1(\%) &RI(\%$\uparrow$) &CSR(\%$\uparrow$)/MCR(\%$\downarrow$)\\ \hline

    %%% \multirow{3}{*}{Original set up} & CodeBERT & 62.74 & 62.31  & 47.81  & 54.11 &- & &81.98 & 82.12 & 81.98& 81.45 &- &  \\ 

    %%% & RoBERTa & 61.56 & 57.71 & 61.11 & 59.36 & - & -& 80.02 & 79.91 & 80.02 & 79.40&-&- \\ 

    %%% & GraphCodeBERT & 62.40 & 61.50 & 48.20 & 54.09 &- &- &81.91 &  81.77& 81.91& 81.56&- & -\\\hline

    \multirow{3}{*}{ CodeImprove-rand} & CodeBERT & 64.86  &64.64&51.87   &57.55  & 12.3  & 11.1/2.16 &82.17  & 82.08&82.18 &81.69  &3.2  &8.6/0.8\\ 

    & RoBERTa & 63.57 & 59.84 & 62.95 & 61.35& 20.7 &6.11.1/1.5 &80.95  &80.85  &80.95 &80.39 &20.9  &17.3/1.36   \\ 

    & GraphCodeBERT& 63.21&62.45  &49.96  &55.51  &9.9  &7.42/0.82 & 82.12&82.18 &82.12&81.75 &2.8  &13.9/1.4  \\\hline

     \multirow{3}{*}{ CodeImprove-HC} & CodeBERT & 63.79  & 63.43 &50.03& 55.94& 6.1 & 6.8/1.9 & 82.22 & 82.24&82.27 &81.79  & 5.03  & 8.3/0.7 \\ 

    & RoBERTa & 62.70 & 58.97 &61.84 &60.37 &11.7 &6.7/0.7 &80.54 &80.44&80.54 &79.98  & 11.8 & 11.9/1.1    \\ 

    & GraphCodeBERT&62.99 &62.33  &49.17  &54.97  &7.2 &5.15/0.5 &82.04 &82.03& 82.03 &81.69 & 1.7 & 10.6/1.1 \\\hline

    \multirow{3}{*}{ CodeImprove} & CodeBERT & \textbf{71.52} 
      & \textbf{70.79}  & \textbf{64.70}&\textbf{67.61}&\textbf{51.28} & \textbf{39.9/4.5} &\textbf{84.5} &\textbf{84.2}&\textbf{84.4} &\textbf{83.99}  & \textbf{44.2}&  \textbf{32.8/1.0}\\ 

    & RoBERTa &\textbf{68.81} &\textbf{64.67}  &\textbf{70.76}  &\textbf{67.58}  &\textbf{74.4} &\textbf{36.2/1.96} &\textbf{82.91} &\textbf{82.99} &\textbf{82.91} &\textbf{82.36} &\textbf{65.6}  & \textbf{32.13/1.0}  \\ 

    & GraphCodeBERT &\textbf{65.26} &\textbf{64.22} &\textbf{55.06} &\textbf{59.29} &\textbf{35.1}  &\textbf{23.1/1.9} &\textbf{83.45} &\textbf{83.55} &\textbf{83.45} &\textbf{83.09} &\textbf{21.1}  &\textbf{27.2/1.4} \\\hline

\end{tabular}}
\end{table*}
\subsection{RQ3: Effectiveness of Search Strategies to Adapt Out-of-Scope Inputs. }

\textit{Baseline:} We employed two search strategies; namely random search (\textbf{CodeImprove-rand})~\cite{zabinsky2009random}, and Hill climbing algorithm (\textbf{CodeImprove-HC})~\cite{selman2006hill}. \textbf{CodeImprove-rand} applies random transformations until identifying the optimal candidate. \textbf{CodeImprove-HC} follows the principles of the hill climbing algorithm.

\textit{Process:} For CodeImprove-rand, transformation operators are applied randomly until the algorithm identifies the best candidate. In CodeImprove-HC, the process starts with an initial solution obtained through a random transformation. The algorithm then iteratively applies a single transformation operator to improve the fitness score. Once the current solution surpasses the fitness threshold, the algorithm terminates, having reached the optimal solution. To ensure fairness, all techniques are limited to 15 transformations per solution. In CodeImprove, each candidate solution undergoes all 15 operators during the crossover phase.

%For CodeImprove-rand, we randomly apply transformation operators until the algorithm finds the best candidate. CodeImprove-HC begins with an initial solution (i.e., obtained through a random transformation). Subsequently, it iteratively applies a single transformation operator to this solution while computing the fitness score. If the current solution surpasses the threshold for fitness score, the algorithm terminates, having achieved the optimal solution. To be fair, we set up a number of transformations for a solution to 15 in all the techniques. For CodeImprove, each candidate solution undergoes all 15 operators during crossover. 

\textit{Results: } Table~\ref{Tab:evaluation} compares the three approaches, highlighting the following key observations:
%Table~\ref{Tab:evaluation} shows the comparison of the three approaches. From this table, we observe that: 
(1) CodeImprove obtained the best accuracy across all subjects (up to 8.78\%) while both CodeImprove-rand and CodeImprove-HC did not achieve the performance of CodeImprove (up to 2.13\%); (2) Although CodeImprove-HC and CodeImprove-rand improve the model performance, we find that these search algorithms stop at the local minima (i.e., once the algorithm identifies a better candidate, the process terminates). However, CodeImprove will evolve for multiple generations i.e., in our case, is three to find the best candidate; (3) In terms of correcting mispredictions CodeImprove performs the best (i.e.,  CSR up to 39.9\%); (4) Although CodeImprove-rand and CodeImprove-HC shows lower values of MCR, note that its CSR values are really low, therefore, unable to correct mispredictions in a large scale; and (5) In conclusion, CodeImprove is a stable approach to adapt program inputs.  

\begin{tcolorbox}[title=\textbf{RQ3} - How effective to convert out-of-scope data to become in-scope data?, left=2pt, right=2pt,top=2pt,bottom=2pt]
CodeImprove is better at correcting out-of-scope inputs compared to other search algorithms such as random search and hill climbing. 
\end{tcolorbox}

\subsection{RQ4: Influence of Hyper-parameters}
\begin{table}[htb!]
\caption{Sensitivity Study}
\label{Tab:sensitivity}
\renewcommand{\arraystretch}{}
\resizebox{\linewidth}{!}{
\begin{tabular}{ccccccc}
\hline
 \multirow{2}{*}{\textbf{Experiment}}  & \multirow{2}{*}{\textbf{Model}} &\multicolumn{2}{r}{\textbf{Vulnerability Detection}} & \multicolumn{2}{r}{\textbf{Defect Prediction}}\\ \cmidrule(lr){3-4} \cmidrule(lr){5-6} %  {3-12}
    &   &RI(\%$\uparrow$)  &CSR(\%$\uparrow$)/MCR(\%$\downarrow$)&  RI(\%$\uparrow$) &CSR(\%$\uparrow$)/MCR(\%$\downarrow$)\\ \hline

 %  {3-12}

    \multirow{3}{*}{$N$=1} & CodeBERT & 48.4 & 37.6/4.32  & 28.1 & 26.1/1.33  \\  

    & RoBERTa & 73.0   &35.1/1.78  & 55.5&29.71/1.25 \\  

    & GraphCodeBERT & 32.9& 21.6/1.8  &18.5 &25.6/1.53 \\\hline

    \multirow{3}{*}{ $N$=5} & CodeBERT & 50.8  & 40.3/5.0 & 30.2 & 27.1/1.31\\  

    & RoBERTa & 71.1  &34.3/1.78 & 56.94 & 30.1/1.23  \\  

    & GraphCodeBERT& 34.2 &  21.4/1.6  &18.95&25.9/1.55\\\hline

     \multirow{3}{*}{$N$=10} & CodeBERT &  51.1 & 38.7/3.83  &42.7    &32.2/1.31 \\  

    & RoBERTa & 71.3   & 35.4/2.14&59.68&30.3/1.12   \\  

    & GraphCodeBERT& 35.1 &22.5/1.8 &19.9&26.6/1.5\\\hline

    \multirow{3}{*}{ $N$=15} & CodeBERT & 51.28
      & 39.9/4.51   &44.2    &32.8/1.0 \\  

    & RoBERTa &  74.4  & 36.2/2.14&65.6& 32.3/1.08  \\  

    & GraphCodeBERT & 35.1   &23.1/1.9 &21.1&27.1/1.4 \\\hline

\end{tabular}}
\end{table}
\textit{Process:} We studies the influence of number of transformation operators ($N$) applied during the crossover for each candidate. We investigated the effectiveness and efficiency of Codeimprove under different settings, i.e., $N$ = \{2, 5, 10, 15\}. We applied variable renaming and one random transformation operator from Table~\ref{table:transformations} for $N=2$, operators 1-5 for $N$=5, operator 1-10 for $N$=10, and operators 1-15 for $N$=15.

\textit{Results: }Table~\ref{Tab:sensitivity} show the results of CodeImprove under the hyper-parameter settings of $N$ in terms
of CSR, MCR, and RI. We observed that as N increases, more mispredicted inputs can be corrected, resulting in a larger RI.
%We found with the setting of $N$ increasing, the more mispredicted inputs can be corrected resulting to a larger RI. 
Meanwhile more correctly predicted inputs are identified as mispredicted ones due to more transformations leading to slightly higher MCR. When $N$ = 5 for CodeBERT on vulnerability detection task, CodeImprove achieved a higher CSR (i.e., 40.3\%) than $N$=10 (i.e., CSR of 38.7\%), however the RI was lower than $N$ =10 due to higher MCR value. Therefore, it is necessary to maintain the balance between CSR and MCR during the transformation. Moreover, CodeImprove has achieved greater performance on all cases for $N$, indicating its practical applicability.

\begin{tcolorbox}[title=\textbf{RQ4} - How sensitive is CodeImprove’s performance under different experimental setups?, left=2pt, right=2pt,top=2pt,bottom=2pt]
CodeImprove is effective when different number of transformation operators are applied during the crossover (i.e., RI shows 48.4\% - 51.28\% for vulnerability detection and 26.1\%-32.8\% for defect detection tasks).
\end{tcolorbox}
\subsection{RQ5: Semantic Preservation in CodeImprove's Program Transformation}

\textit{Process: } The objective of this RQ is to examine whether the adapted programs maintain the semantics of the original inputs. We investigate the effectiveness of applying semantic preserving program transformations to adapt out-of-scope inputs. Based on our investigation we provide an example in Figure~\ref{fig-mc}. Additional examples are on our project website due to space restrictions\cite{CodeImprove}.

%we will add more examples to our project website\textcolor{blue}{\cite{CodeImprove}}. 

Figure~\ref{fig-mc} illustrates how CodeImprove revises a misprediction. As illustrated in Figure~\ref{fig-mc-ori}, the CodeBERT model incorrectly predicts the input to \textit{no defect}, although the ground truth label is actually \textit{wrong output}. During the validation phase, CodeImprove identifies this out-of-scope input with a validity score of 0.0221, significantly lower than our threshold of 0.2. To adapt this input, CodeImprove applies semantic preserving transformations. These transformations include splitting lines, changing code order,  splitting declarations, and separating variable assignments at line 6. Additionally, the relational and incremental operators were altered in lines 9 and 12. These changes create a syntax shift that affects the model's interpretation, resulting in different embeddings. The transformed version is shown in Figure~\ref{fig-mc-transed}. After these transformations, the model correctly predicts the label as \textit{wrong output} with an improved validity score of 0.7377. 

%including splitting the single-line variable definition into the multiple-line definition at line 6, reversing both the operators and operands at lines 9 and 12, and replacing the self-increment operator into a variable assignment with an algebraic expression at lines 9 and 12, resulting in the transformed version shown in Figure~\ref{fig-mc-transed}. After applying these transformations, the model correctly predicted the label to \textit{wrong output} with an improved validity score of 0.7377. 

\begin{figure}[t]
  \centering
  \subcaptionbox{Before Transformation  \label{fig-mc-ori}}{
\includegraphics[width=0.2\textwidth]{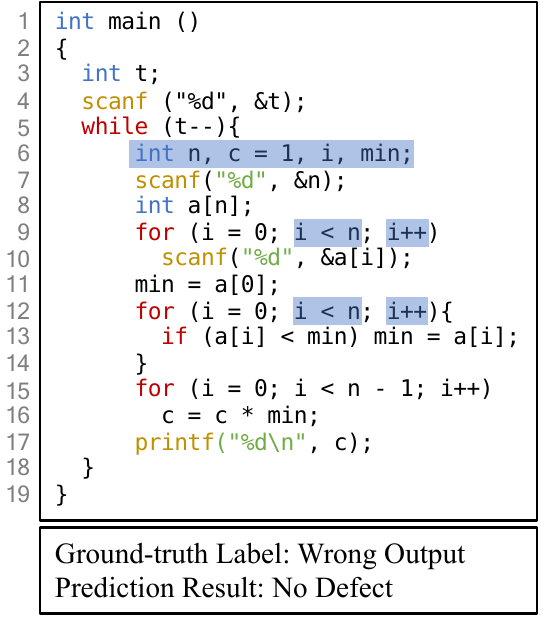}
  }
  \subcaptionbox{After Transformation\label{fig-mc-transed}}{
    \includegraphics[width=0.22\textwidth]{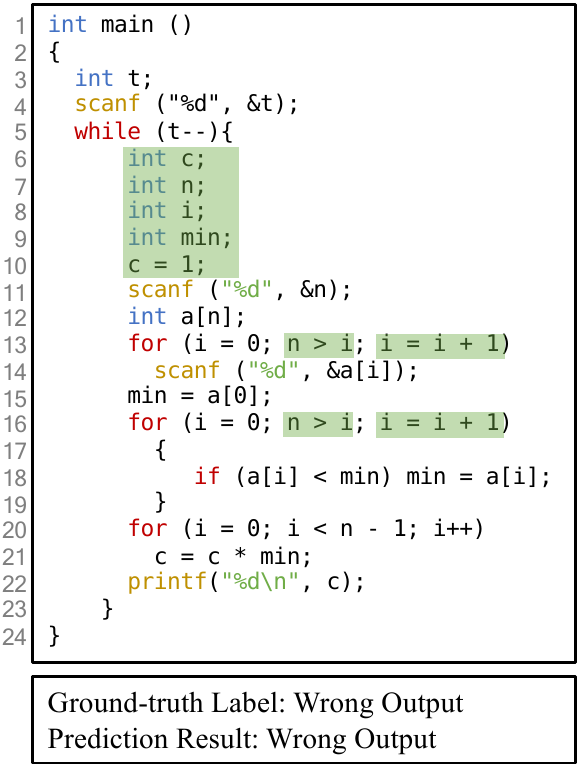}
    }
  \caption{An example of a M$\rightarrow$C transformation}
  \label{fig-mc}
\end{figure}

\begin{tcolorbox}[title=\textbf{RQ5} - Does program transformations of
CodeImprove preserve semantics?, left=2pt, right=2pt,top=2pt,bottom=2pt]
CodeImprove can generate semantic preserving program transformations. 
\end{tcolorbox}

\subsection{RQ6: Overhead of CodeImprove}

%\textcolor{red}{organize this section to show online offline stage.}
\textit{Process: } To apply CodeImprove in real-time, we compute the overhead of applying transformations for an input. We calculate the TPS, which measures the number of transformations applied per second by CodeImprove. This metric is averaged across all $N$ variants of CodeImprove studied under RQ4. Additionally, we evaluate both offline and online overhead. The offline stage involves the sub-model training, while the online stage measures the time required to adapt an input with CodeImprove. Table~\ref{Tab:rq5} shows the statistics of TPS, and Figure~\ref{fig:combined_figures} shows the time overhead of CodeImprove.

%\textit{Process: }  In order to apply CodeImprove in real-time, we compute the overhead of applying transformation for an input. We compute the TPS, which measures the number of transformations applied per second by CodeImprove.  This is the average of all $N$ variants of CodeImprove studied under RQ4. Table~\ref{Tab:rq5} shows the statistics of overhead for adapting an out-of-scope input by CodeImprove. %Note that we applied CodeImprove on all samples in Table~\ref{newstats}. 

\begin{table}[h!]
\caption{TPS of Codeimprove}
\label{Tab:rq5}
\renewcommand{\arraystretch}{1}
\resizebox{\linewidth}{!}{
\begin{tabular}{| c | c | c | c | c | c | c | c |}
\hline
 \multirow{2}{*}{\textbf{Experiment}} &\multicolumn{3}{c|}{\textbf{Vulnerability Detection}} & \multicolumn{3}{c|}{\textbf{Defect Prediction}}\\ \cline{2-7}
      & CodeBERT & RoBERTa & GraphCodeBERT  & CodeBERT & RoBERTa & GraphCodeBERT \\ \hline

    Overhead  & 1.51 & 1.2  & 1.43  & 2.04 & 1.95 & 1.75  \\\hline
  
\end{tabular}
}
\end{table}

%\vspace{-1em}
Based on  Table~\ref{Tab:rq5}, CodeImprove is capable of applying transformations at a rate of 1.2 TPS to 2.04 TPS for each SE task across all code models. Figure~\ref{fig:figure2} confirms that CodeImprove takes approximately 49.92s to 59.4s to adapt an input across all models. We plan to further minimize the transformation times in future work.  Moreover, CodeImprove is more efficient and offers a practical, scalable solution to enhance model performance without the significant cost and time investment required by traditional methods such as retraining and replacement.
%\vspace{-1em}

%\vspace{-1em}
It is important to note that the sub-model training procedure is treated as an offline stage, minimizing its impact on overall performance. Figure~\ref{fig:figure1} shows that training a sub-model takes around 900s to 940s for the vulnerability detection  and 1200s to 1250s for the defect prediction across all models on a machine with an NVIDIA GeForce GTX 1080 GPU. This process only needs to be done once and incurs significantly lower costs compared to regular retraining or fine-tuning.

\begin{figure}[H]
    \centering
    \begin{subfigure}[b]{0.49\columnwidth}
        \centering
        \includegraphics[width=\columnwidth]{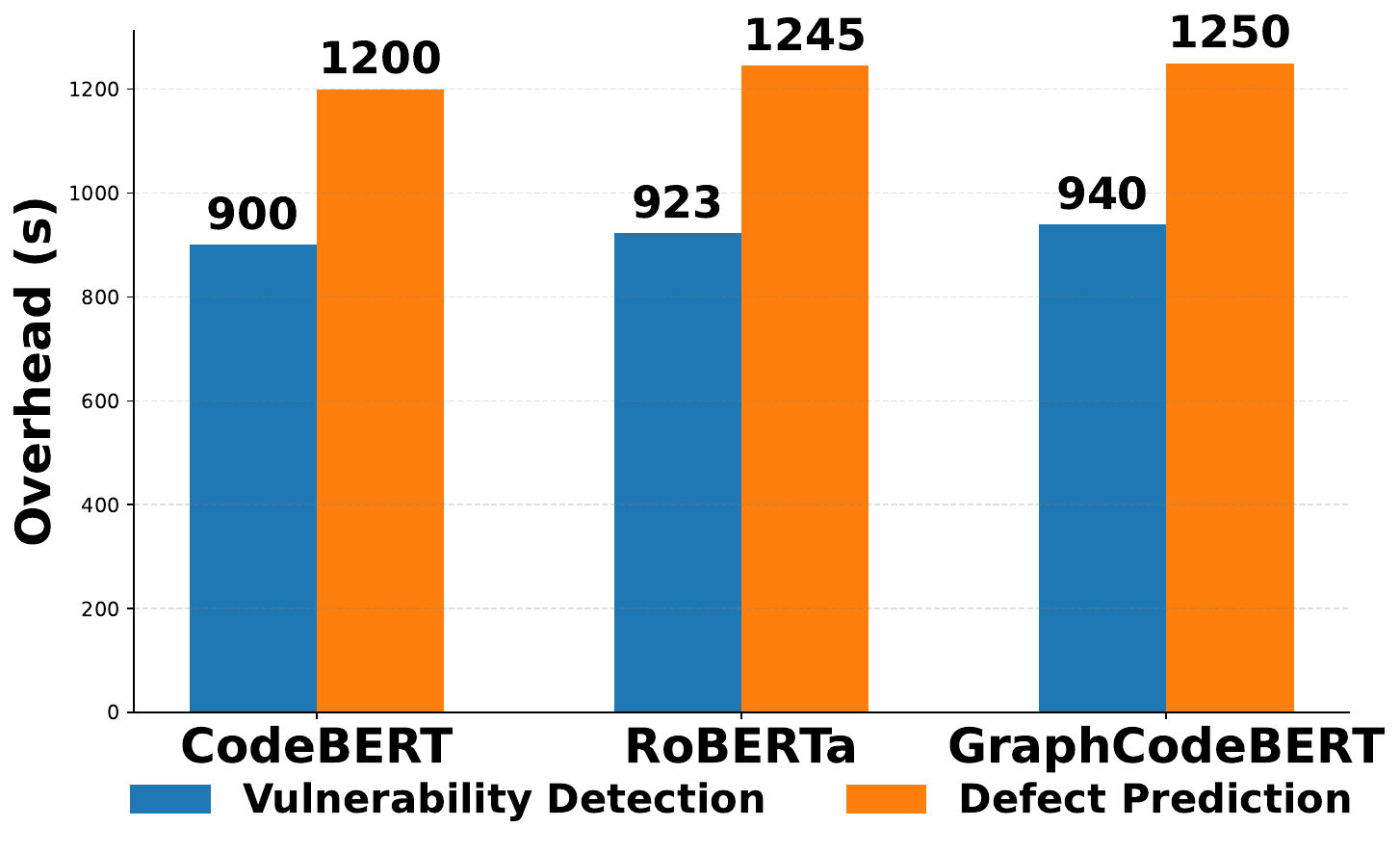}
        \caption{Offline time overhead (in second)}
        \label{fig:figure1}
    \end{subfigure}
    \hfill
    \begin{subfigure}[b]{0.49\columnwidth}
        \centering
        \includegraphics[width=\columnwidth]{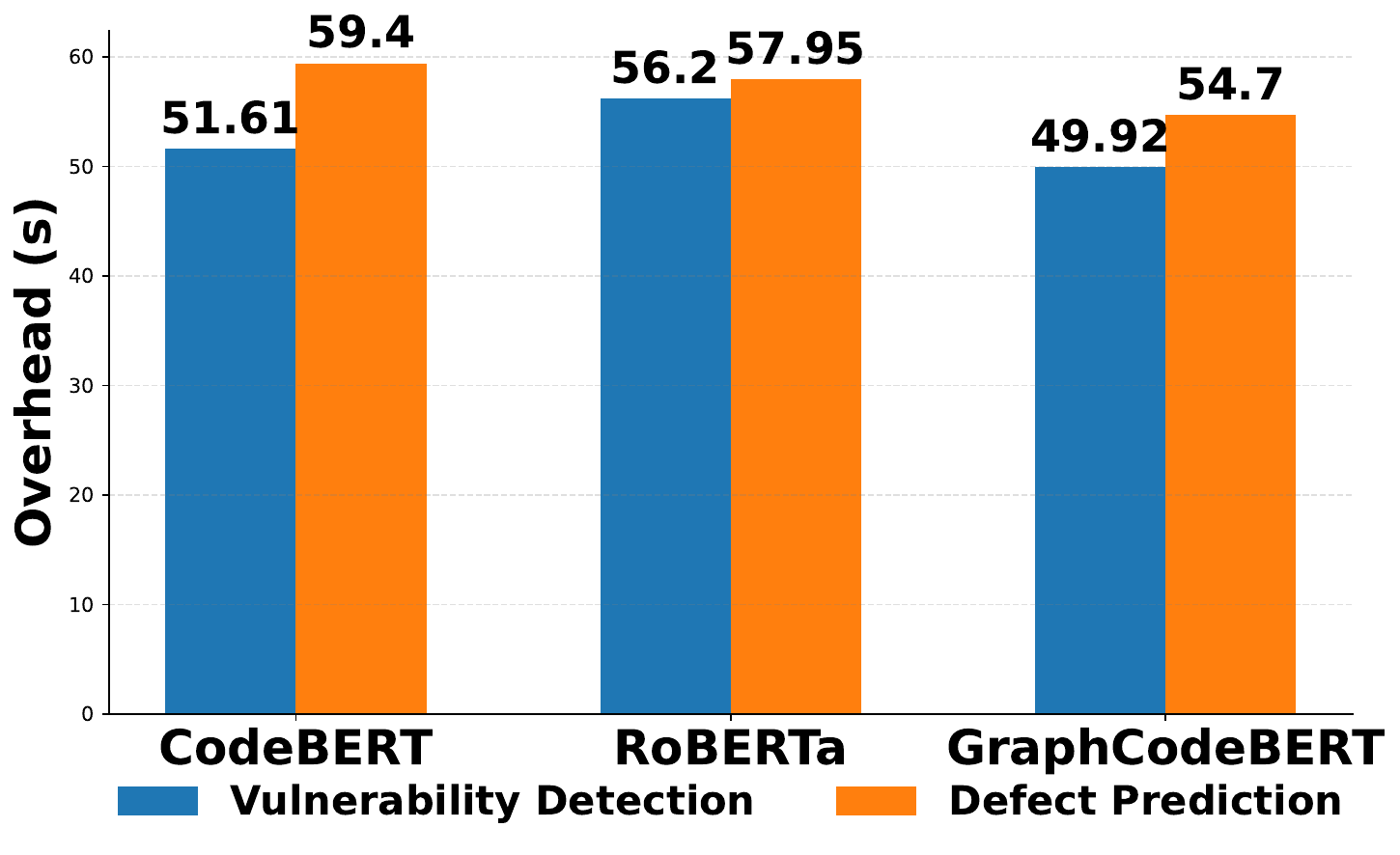}
        \caption{Online time overhead (in seconds)}
        \label{fig:figure2}
    \end{subfigure}
    \caption{Time Overhead of CodeImprove}
    \label{fig:combined_figures}
\end{figure}

\begin{tcolorbox}[title=\textbf{RQ6} - What is the overhead of CodeImprove in adapting a program to DL models?, left=2pt, right=2pt,top=2pt,bottom=2pt]
CodeImprove was highly efficient in adapting an out-of-scope input through semantic preserving program transformations in real-time (1.2TPS - 2.04TPS).
\end{tcolorbox}

%In recent years, lots of researches about uncertainty measurement for DL models have been proposed. We select a subset of uncertainty metrics from the existing literature for their prevalence, scalability, and practical applicability. The selected work includes:

%\input{tex/limitation}

\section{Threat Analyses and limitations}
\label{threat}

Our selection of the two subject datasets, namely, Devign, and CodeChef with their associated code models, might threaten the external validity of our experimental conclusions. We tried to alleviate this threat by following efforts: 
(1) the two datasets are very popular and have been widely used in relevant research~\cite{lu2021codexglue,tian2023code,yang2022natural,carrot}; (2) their associated DL models are commonly used in SE tasks; (3) these datasets and models differ from each other by varying topics, %(e.g., BERT mainly focuses on NL data while codeBERT focuses on both NL-PL data), 
labels (from two to four), and model accuracies (from 61.56\% to 81.98\%), which make these subjects diverse and representative. Therefore, our experimental conclusion should generally hold, although specific data could be inevitably different for other subject.

Threats to external validity may arise from the techniques selected for experimental comparisons, including uncertainty metrics~\cite{guo2017calibration,wang2020dissector,hendrycks2018baseline,gal2016dropout, alon2019code2vec,xiao2019quantifying,vasudevan2019towards,corbiere2019addressing, monarch2021human, steinhardt2016unsupervised,shannon1948mathematical} and dissector~\cite{wang2020dissector}. Due to inherent differences, existing methods for image data (SelfChecker~\cite{xiao2021selfchecking} and InputReflector~\cite{xiao2022repairing}) cannot be directly applied to code data. Therefore, we chose dissector as a baseline for input validation, as it identifies out-of-scope inputs similarly to our focus. Additionally, we sampled a subset of uncertainty metrics to assess their effectiveness in identifying out-of-scope inputs ~\cite{guo2017calibration,wang2020dissector,hendrycks2018baseline,gal2016dropout, alon2019code2vec,xiao2019quantifying,vasudevan2019towards,corbiere2019addressing, monarch2021human, steinhardt2016unsupervised,shannon1948mathematical}.

Our internal threat mainly comes from the lack of ground truths for distinguishing out-of-scope inputs from in-scope inputs, limitations on applying CodeImprove in different subjects, and applying different program transformation rules. We used mispredictions and correct predictions to simulate out-of-scope inputs and in-scope inputs respectively. Such estimation can be rough, however the logic may holds (RQ2). Regarding the lack of subject matter, our assumption was to investigate how our propose technique performed on different SE tasks and code models. Therefore, we select a subset of SE tasks and code models (RQ1, RQ3 and RQ4). In future, we plan to extend and apply CodeImprove on other subjects matter (e.g. clone detection, functionality classification etc.). CodeImprove uses 15 semantic-preserving transformation rules, but we plan to add more and update our website with new results. Due to the diversity of our subjects, we believe our conclusions are generally applicable, and CodeImprove can support future research.

%Currently, CodeImprove employs 15 semantic preserving program transformation rules. However, in future, we plan to add more transformation rules and update our project website with the latest results.
%Therefore, considering our subjects have necessary diversities as discussed above, our experimental conclusions can hold generally and we believe that the technique applied in CodeImprove can be used for future research.

%We believe that the technique applied in CodeImprove can be used for future research.% however, in future, we plan to add more transformation rules and update our project website with the latest results. 
%Therefore, considering our subjects have necessary diversities as discussed above, our experimental conclusions can hold generally and we believe that the technique applied in CodeImprove can be used for future research. %that can help to confront the code data shifts and improve our genetic based algorithm to manage large search space. Therefore, considering our subjects have necessary diversities as discussed above, our experimental conclusions can hold generally and we believe that the technique applied in CodeImprove can be used for future research. 

\vspace{-1em}

\section{Related Work}
\label{related}
%Our work in this paper provides a method to improve deep code models' ability to deal with shifted code data without retraining deployed models, through program transformation techniques. Therefore, this section will briefly introduce some recent representative work in these fields, including code transformation, input validation and input adaptation. 
%\vspace{-1em}
\textbf{Input Validation for Deep Learning Model.} %As deep learning models are deployed in many critical scenarios, several works have been proposed to check if the model can correctly process given input, i.e. compute the trustworthiness of a result predicted by a deployed DNN.
Several work has been proposed to compute the trustworthiness of a DL model. Hendrycks et. al~\cite{hendrycks2018baseline} propose a baseline for detecting misclassified samples. \cite{guo2017calibration} propose re-calibration of probabilities on a held-out validation set. \cite{wang2020dissector} proposed Dissector, to validate inputs by crossing-layer dissection. 
\cite{xiao2021selfchecking} proposed SelfChecker by leveraging Kernel Density Estimation(KDE)\cite{terrell1992variable}. ConfidNet\cite{corbiere2019addressing} is designed to learn the confidence criterion using True Class Probability for predicting failures. InputReflector \cite{xiao2022repairing} identifies failure-inducing inputs on Image data. In addition to the aforementioned techniques on code data, we show that existing uncertainty metrics~\cite{guo2017calibration,wang2020dissector,hendrycks2018baseline, li2021estimating, alon2019code2vec, vasudevan2019towards,corbiere2019addressing} do not perform promising results on code data.

\textbf{Related Work on Code Inputs.} 
Several techniques\cite{zhang2020generating,yefet2020adversarial, naturalattack, carrot,srikant2021generating} for generating adversarial code to challenge these models have been proposed in recent years. Tian et al.\cite{tian2023code} claim they designed a code-difference-guided generation technique, which can improve the efficiency further. CodeDenoise\cite{tian2023fly} is the most advanced technique to improve the performance of deployed models without retraining, however, it can only relieve the noise introduced by different identifier names and consequential mispredictions. In contrast, CodeImprove leverages 15 unique transformation operators. To the best of our knowledge, CodeImprove is the first attempt to use the inference-time program transformation technique to enhance the performance of code models

\section{Conclusion}
\label{conclusion}
\vspace{-1mm}
This paper proposes CodeImprove for validation and adaptation of out-of-scope inputs to become in-scope inputs for code models. CodeImprove employs a validity score metric by leveraging dropout based sub-model generation (DSMG) technique to distinguish out-of-scope inputs from in-scope inputs and applies semantic preserving program transformations on these inputs. Experimental evaluation confirmed that CodeImprove's effectiveness, highlighting its potential for broader application scenarios in input validation and adaptation. 

\section*{Data Availability}
\label{data}
Our artifacts are available at~\cite{Data}, containing datasets, code to reproduce the results in this paper. Our project website is available at ~\cite{CodeImprove}

\section*{ACKNOWLEDGMENTS}
This work was partially supported by NSF grants NSF CCF-2146443 and Amazon Trust AI Research Award. We extend our gratitude to Mirazul Haque for hiss guidance throughout this project.

%%%oopsla
%We confirm that the source code of the experiments is available at an anonymous GitHub repository for peer review. The GitHub repository includes all source codes to run the experiments and scripts to download the datasets and models for producing the experiment results. The GitHub repository is available at: \href{https://github.com/CodeImprove/CodeImprove/tree/main}{https://github.com/CodeImprove/CodeImprove/tree/main}
%This paper introduces CodeImprove, a method for validating and adapting out-of-scope inputs to in-scope inputs for code models. CodeImprove employs a validity score metric to distinguish between the two types of inputs and applies semantic-preserving program transformations. Experimental results validate CodeImprove's effectiveness, highlighting its potential for broader application scenarios in input validation and adaptation.

%\input{tex/dataavailability}

%% Acknowledgments

%\clearpage
\bibliographystyle{IEEEtran}
\bibliography{conference_101719}
\end{document}